\documentclass[superscriptaddress,prx,longbibliography,twocolumn, footinbib]{revtex4-2}

\usepackage[dvipsnames]{xcolor}
\usepackage[space]{grffile}
\usepackage{amsmath, amssymb, amsfonts,bm}

\usepackage{latexsym}   
\usepackage{bbm} 
\usepackage{mathtools}  
\usepackage{placeins}   
\usepackage{comment}
\usepackage{float}
\usepackage[colorlinks=true, citecolor=blue, linkcolor=blue, urlcolor=blue]{hyperref}
\usepackage{xr-hyper}
\usepackage{bm}
\usepackage[printonlyused]{acronym}
\usepackage{layouts}
\usepackage{url}

\newcommand{\ETH}{Institute for Theoretical Physics, Wolfgang Pauli Str. 27, ETH Zurich, 8093 Zurich, Switzerland}

\newcommand{\Hamb}{Max Planck Institute for the Structure and Dynamics of Matter, Luruper Chausse 149, 22761 Hamburg, Germany}

\begin{document}


\title{Terahertz amplification and lasing in pump-probe experiments with hyperbolic polaritons in h-BN}

\author{Khachatur G. Nazaryan} 
\affiliation{Department of Physics, Massachusetts Institute of Technology, Cambridge, MA 02139}
\author{Ivan Ridkokasha} 
\affiliation{Instituut-Lorentz, Universiteit Leiden, P.O. Box 9506, 2300 RA Leiden, The Netherlands} 
\author{Marios~H.~Michael} 
\affiliation{\Hamb}
\author{Eugene~Demler}
\affiliation{\ETH}
\date{\today}

\begin{abstract}
We discuss terahertz pump-probe experiments in hBN from the perspective of Floquet optical materials and photonic time crystals. Anisotropic nature of this material results in a large separation of frequencies of in-plane and out-of-plane optical phonons and leads to several branches of phonon-polariton excitations. We consider a slab of finite thickness pumped at the frequency of middle polaritons, around 25 THz. We theoretically analyze properties of the pump induced state and identify several interesting features, including possible light amplification and parametric instability at 2 THz. We relate these features to resonant pair scattering processes of polaritons excited by the pump pulse. We show that amplification/instability frequency ranges depend on both the pump intensity and the slab thickness. We  point out that this setup can be used utilized to strongly enhance optical nonlinearities at terahertz frequencies. Richness and tunability of this system suggests that it holds promise for practical applications in terahertz technology.  
\end{abstract}

\maketitle

\section{Introduction}

\begin{figure}[t]
\begin{centering}
\includegraphics[width=0.9\columnwidth]{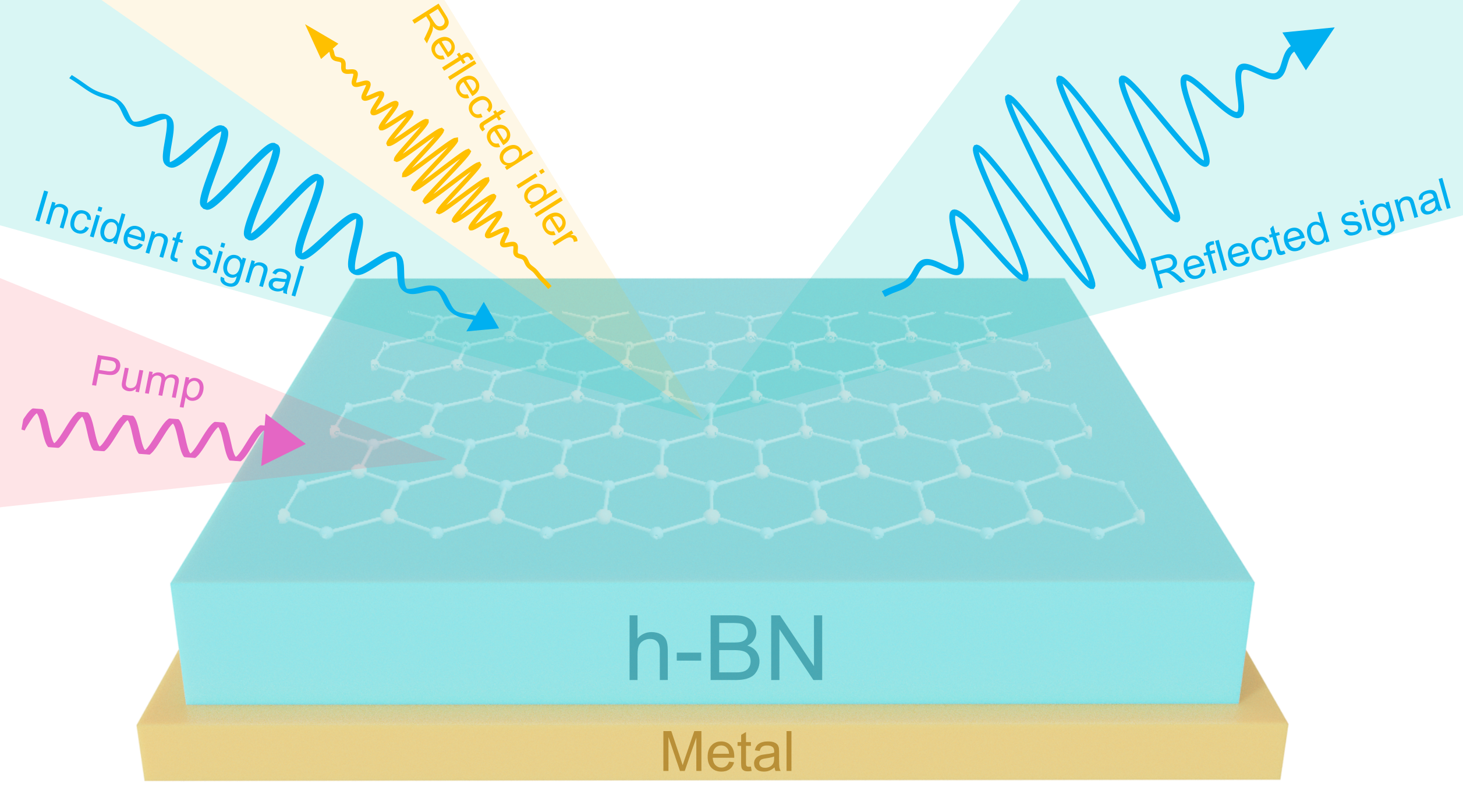}
\par\end{centering}
\caption{Schematic illustration of pumped hBN atop a metallic cladding. 
When hBN is driven at the out-of-plane phonon frequency, $\omega_\text{p} = 23.3$ THz, phonon oscillations transform hBN into a Floquet material / Photonic time crystal. Such a system exhibits lasing instabilities at parametrically resonant frequencies, that come in pairs as signal frequency $\omega_\text{s}$ and idler $\omega_\text{i} = 2 \omega_\text{p} - \omega_\text{s}$. When the system is probed at resonant signal frequencies, emission is amplified by hBN, while simultaneously generating radiation at the idler frequency. In this way pumped hBN can act both as an amplifier and a source of THz radiation.} 
\label{fig:Setup}
\end{figure}

Light propagation in time varying media has recently emerged as a new approach for manipulating photons and controlling light-matter interaction. Systems with periodic temporal modulation are referred to as photonic time crystals (PTC) \cite{Segev_review, Alu_review} or Floquet Optical Materials (FOM) \cite{Marios2022,Sho_22,Marios_thesis, Weiner2011, Cartella2018, Marios_20}. These systems exhibit a variety of surprising phenomena including nonreciprocity of wave propagation \cite{Wang_18}, light amplification \cite{Marios_20, vonHoegen_22}, lasing \cite{Lyubarov22} as well as generation of squeezed collective modes \cite{taherian2024squeezed, Rubaiat23, michael2023theory, Dolgirev22}. Observing FOM phenomena requires changing optical
parameters on time scales comparable to the period of the electromagnetic wave itself. Pump probe experiments in solids provide a natural platform for investigating this class of phenomena. The pump pulse launches oscillations of a collective
mode in the material, which provides temporal modulation of refractive index for the probe pulse. Collective excitations in solids, including phonons and plasmons, typically have terahertz and IR frequencies. When the probe pulse has similar frequency, conditions for observing FOM phenomena are fulfilled.

Theoretical interpretation of FOM phenomena in pump probe experiments in solids require
several additional considerations. Firstly, one needs to account for
the scattering geometry. In these experiments, the probe pulse is incident from and is
reflected to air, which is a stationary medium. Only the transmitted part of the pulse propagates in the Floquet medium of the photoexcited solid. The reflected part of the probe pulse, which is commonly used for detection, acquires signatures of Floquet light dynamics only through boundary conditions at the interface \cite{Marios2022}.  Secondly, the complexity of the material's internal structure plays a significant role. Many materials, possess multiple branches of polaritons -- collective modes originating from the coupling of photons with excitations in solids (e.g. phonons, plasmons and excitons).  For example, in layered materials, different frequencies of in-plane and out-of-plane phonons lead to an anisotropy in the dielectric tensor and result in the emergence of two distinct families of phonon-polaritons. These multiple polariton branches make it possible to realize a variety of nonlinear optical processes, which adds richness to FOM phenomena.

\begin{figure}[t]
\begin{centering}
\includegraphics[width=0.99\columnwidth]{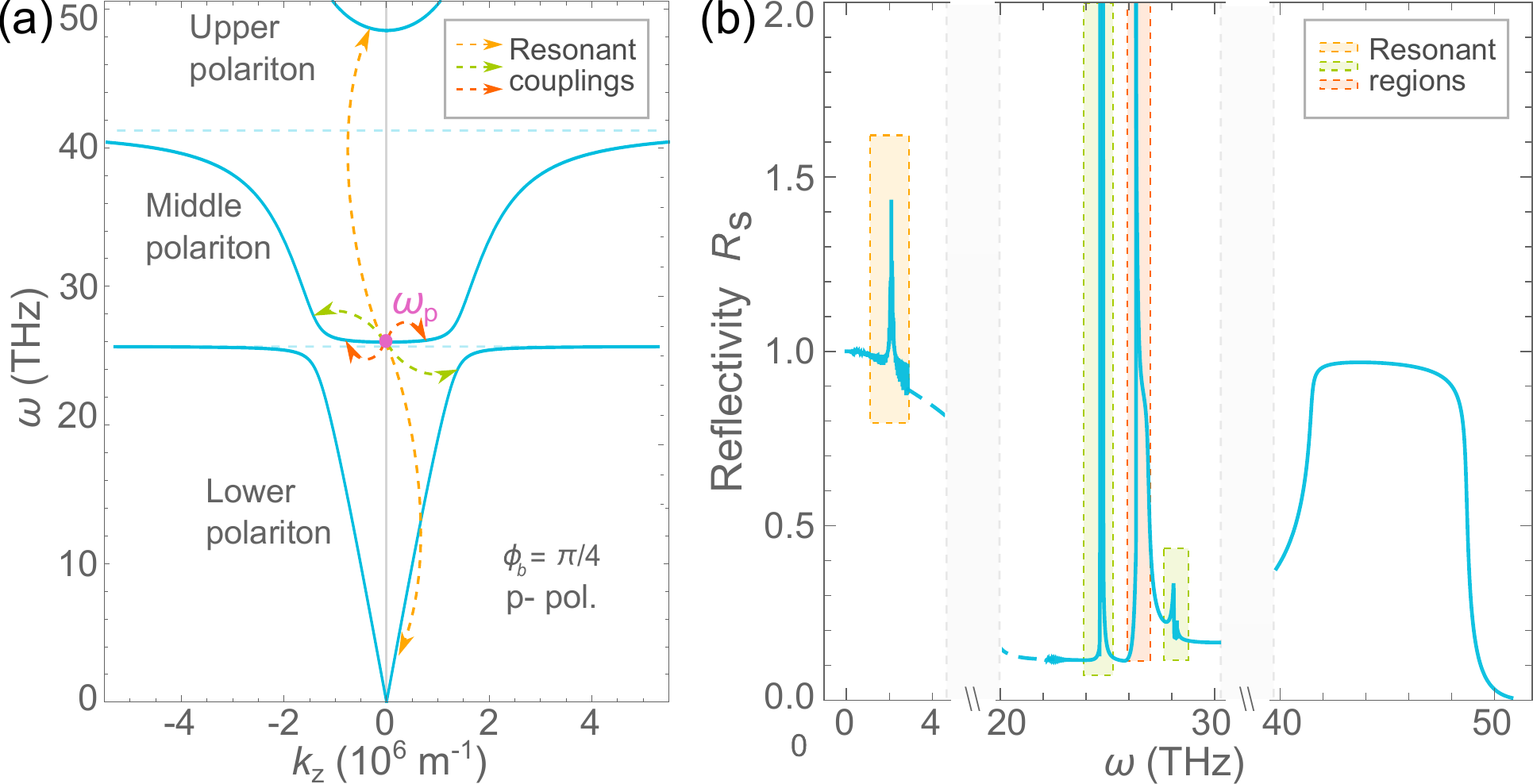}
\par\end{centering}
\caption{(a) Schematic illustration of Floquet parametric instabilities due to resonant polariton scattering processes. We plot the polariton dispersion in hBN arising from 
the $41.1$ THz in-plane and $23.3$ THz out of plane TO phonons, (see section~\ref{Sec:Hyperb Phon-Pol}). In the proposed scheme, the pump pulse resonantly excites the optical phonon at $23.3$ THz (pink circle). Non-linearities lead to the breakup of the coherent phonon oscillations into polariton pairs at signal and idler frequencies which conserve energy $2\omega_\text{p}=\omega_\text{s}+\omega_\text{i}$ and momentum $\textbf{k}_\text{s}=-\textbf{k}_\text{i}$. This process leads to parametric instabilities annotated by the colored arrows: yellow for the resonant coupling of the lower and upper band, green for the lower and middle band, and red for middle band instabilities near $\omega_\text{p}$. 
(b) Reflectivity versus signal frequency for s- and p- polarized light in a finite slab. Each resonant process is accompanied by parametrically amplified reflectivity which can exceed unity. The different resonant regions are highlighted with colors matching the corresponding resonant process. Details are discussed in Sections \ref{Sec:2THz}, \ref{Sec:48.5THz}, and \ref{sec:Other}.
}
\label{fig:Resonant couplings}
\end{figure}

In this paper, we delve into the exploration of light-matter interaction and parametric amplification in layered anisotropic materials. We extend the previous analyses \cite{Sho_22,Marios2022} by exploring the rich tapestry of anisotropic phonon polaritons in these systems. For concreteness, the primary subject of our discussion  
is phonon-polaritons in hexagonal boron nitride (hBN), although we expect our formalism and results to be applicable to generic anisotropic materials.  hBN, with its two-dimensional structure of alternating boron and nitrogen atoms arranged in a hexagonal lattice, exhibits several properties that make it a good candidate for our study \cite{Watanabe2004, Kim2020,Giles2017, CaldwellTHz2014, Geim2013}. Its insulating nature \cite{Laturia2018}, coupled with its wide bandgap  \cite{Watanabe2004,Cassabois2016}, high thermal conductivity \cite{Zheng2016}, and robust mechanical strength \cite{Ding2016}, make it ubiquitous for experimental studies and suitable for various technological applications \cite{Caldwell2014, Zomer2011, Wang2022, Kretinin2014, Dean2010, Tran2016, Shotan2016}.  

A layered structure of hBN results in very different frequencies for in-plane and out-of-plane transverse optical (TO) phonons.  Their values are  41.1 THz (in-plane mode) and 23.3 THz (out-of-plane mode) respectively. Away from high symmetry directions, longitudinal optical (LO) and TO phonon-polaritons hybridize, giving rise to an intricate dispersion of multiple polariton modes. The presence of different branches leads to several possibilities of resonant scattering of photo-excited polaritons, which will be crucial for our analysis. We note that the strong optical anisotropy of hBN has also attracted interest in the context of hyperbolic phonon polaritons (HPPs), which are expected to feature strong light matter interaction \cite{Dai2014, Giles2017, Caldwell2014, CaldwellTHz2014, Pakdel2014,Ashida23}. This property can be utilized to enhance nonlinear interactions between modes. However, in this paper we will only consider intrinsic nonlinearities of phonon-polaritons in hBN.

The system we consider consists of an hBN slab of finite thickness placed atop a metallic cladding, which acts as an optical mirror. We consider a pump pulse that excites out-of-plane phonon polaritons at a frequency of approximately 25.3 THz and zero momentum. Coherent oscillations of these polaritons create an effective FOM \cite{Weiner2011, Cartella2018, Sho_22, Marios2022}. Another perspective on processes that will be central to our discussion is that of {\it stimulated resonant scattering of polariton pairs}. The pump pulse creates a large population of polaritons at the bottom of the middle polariton branch shown in Figure \ref{fig:Resonant couplings}(a). Due to phonon nonlinearities, pairs of photoexcited polaritons can undergo scattering processes into different types of polaritons, provided that scattering processes satisfy energy conservation and phase matching conditions \cite{textbook_Yariv}.
A probe pulse at frequency $\omega_\text{s}$ (signal frequency), enhances pair scattering processes in which frequency of one of the photons matches the probe photons  and the other photon is at the complementary idler frequency $\omega_\text{i}$ (see eq (\ref{omega_i})). Such resonant processes lead to renormalization of reflectivity, including amplification, and even lasing. 

Our analysis employs the following strategy. We use Maxwell's equations in conjuction  with semiclassical equations of motion of phonon quadratures to derive linearized equations describing two-mode mixing in the Floquet medium (signal and idler frequency components). Signal and idler components of a Floquet eigenmode have momenta that are opposite of each other, and their frequencies add up to the modulation frequency of the index of refraction of the system. The latter corresponds to twice the frequency of the pump photons $\omega_p$.  This classical phase matching picture corresponds to the microscopic interpretation of two pump polaritons at frequency $\omega_p$ and zero momentum resonantly scattering into a pair of polaritons with momenta and energies $\textbf{k}_\text{s}$, $\omega_\text{s}$ and $\textbf{k}_\text{i}$, $\omega_\text{i}$,
\begin{eqnarray}
    2\omega_\text{p}= \omega_\text{s}+\omega_\text{i}, \quad \textbf{k}_\text{s}=-\textbf{k}_\text{i}.
    \label{omega_i}
\end{eqnarray} 
 Linearized light-matter equations give us Floquet eigenfrequencies and eigenmodes of the pumped system \cite{Sho_22,Marios2022}. In the case of strong modulation (this requires large intensity of the pump pulse and substantial nonlinearities), imaginary part of Floquet eigenfrequencies can become positive, indicating parametric instability. We use analysis of Floquet eigenmodes to solve the Fresnel-Floquet light reflection problem \cite{Marios2022} and compute the reflection coefficient of the system in Figure \ref{fig:Setup}. We find that even before the system becomes unstable, reflectivity can be much larger than unity (see Fig. \ref{fig:Resonant couplings}(b)), which indicates that the system becomes a gain medium. 

One of the propitious
 results of our analysis is the possibility of achieving gain at frequencies around 2 THz. The gain frequency range can be modified by varying the intensity of the pump pulse. This can be understood as a result of nonlinear shifts of polariton frequencies and enhancement of parametric instabilities with increasing pump amplitude. Additional tunability of the gain regime can be achieved through varying geometry of the hBN slab.
These results suggest that the system we discuss in this paper holds promise for applications as tunable THz lasers and amplifiers \cite{THz_gap}.


Before concluding this section, it is useful to note that FOM
phenomena, which we discuss in this paper, exhibit some conceptual similarities to physics of
optical phase conjugation \cite{Zeldovich_book}. One
important difference is that terahertz pump probe experiments
utilize electromagnetic waves with wavelengths that are much
larger than the lengthscale over which optical parameters vary.
Thus analysis of light reflection in our case requires going beyond
the Slowly Varying Envelope Approximation used in the
discussion of OPC at optical frequencies. Another class of related
systems is resonant parametric scattering of exciton-polaritons in
semiconductors \cite{Keeling_Berloff_review}.

\section{Nonlinear model of phonon-polaritons in hBN}



\subsection{Equations of motion}

In hBN infrared (IR) active phonons give rise to two branches of transverse optical phonons: $\Omega_{\text{TO},\parallel}$ for the in-plane direction of electric polarization (along the hBN plane), and $\Omega_{\text{TO,z}}$ for the perpendicular direction of polarization \cite{Kumar2015}. 
We denote the phonon displacements as $\textbf{Q}_\parallel$ and $\textbf{Q}_z$, for the in-plane and out-of-plane directions respectively. For convenience, we  rescale $\textbf{Q}$ to $\textbf{Q}/\sqrt{M}$, where $M=M_B M_N /(M_B+M_N)$ is the reduced mass of the unit cell. Combined equations of motion for the electric field and phonon displacements can be written as (see sections \ref{SecSup:Hamiltonian}, \ref{SecSup:EqMot} of the supplementary material for details)
\begin{eqnarray}
    &\ddot{\textbf{Q}}_\beta +\gamma_\beta \dot{\textbf{Q}}_\beta= -\tilde{\Omega}^{2}_\beta \textbf{Q}_\beta+\tilde{\eta}_\beta \textbf{E}_\beta, \label{Q_EOM},\\
    &\nabla_\beta \left(\nabla  \textbf{E} \right)-\nabla ^2 \textbf{E}_\beta +\frac{1}{c^2}\partial_t ^2 \left( \epsilon_{\infty,\beta} \textbf{E}_\beta +\tilde{\eta}_\beta \textbf{Q}_\beta\right)=0 \label{E_EOM}.
\end{eqnarray}
Here, index $\beta=\parallel,z$ denotes the in-plane and out-of-plane directions respectively, $\gamma_\beta$ describes the damping of phonons, $\textbf{E}$ stands for the electric field and $c$ is the speed of light in vacuum. Additionally, $\tilde{\Omega}^{2}$, and effective charge $\tilde{\eta}$ are expressed as,
\begin{eqnarray}
    &\tilde{\Omega}^{2}_{\beta} \left(\textbf{Q},\textbf{E}\right) = {\Omega^{2}_\beta - 4\zeta Q^{2}} +2\alpha\left(\textbf{Q}\textbf{E}\right), 
     \label{nonlinear_terms1} \\
    & \tilde{\eta}_{\text{TO},\beta} \left(\textbf{Q},\textbf{E}\right) = \eta_\beta +\alpha Q^{2}.
    \label{nonlinear_terms2}
\end{eqnarray}
The last two equations describe nonlinear interactions between modes through renormalization of the phonon frequencies, $\tilde{\Omega}^2_\beta$ and their effective charge, $\tilde{\eta}_{\text{TO},\beta}$. The parameters $\alpha, \zeta$ control the non-linear interactions. For simplicity, we assume rotationally invariant form of nonlinearities. Table \ref{tab:Parameters} presents microscopic parameters of the system as reported in Ref. \cite{Iyikanat2021}.

\subsection{Phonon-polaritons in the bulk material \label{Sec:Hyperb Phon-Pol}}

\begin{table}[!t]
\begin{centering}
\begin{tabular}{|c|c|c|c|}
\hline 
$\Omega_{\text{TO},\parallel}$ & $41.1$ (THz) & $\eta_{\parallel}$ & $1.05\times10^{9}$ $(\text{A.s}/\text{kg}^{1/2}\text{m}^{3/2})$\tabularnewline
\hline 
$\Omega_{\text{TO},z}$ & $23.3$ (THz) & $\eta_z$ & $2.73\times10^{8}$ $(\text{A.s}/\text{kg}^{1/2}\text{m}^{3/2})$\tabularnewline
\hline 
$\gamma_\parallel$ & $0.15$ (THz) & $\alpha$ & $1.9\times10^{26}$ $(\text{A.s}/\text{kg}^{3/2}\text{m}^{1/2})$\tabularnewline
\hline 
$\gamma_z$ & $0.12$ (THz) & $\zeta$ & $5.2\times10^{45}$ $(\text{A.s}.\text{V}/\text{kg}^{2}\text{m})$\tabularnewline
\hline 
$\epsilon_{\infty,\parallel}$ & $4.87$ & $\epsilon_{\infty,z}$ & $2.95$ \tabularnewline
\hline 
\end{tabular}
\par\end{centering}
\caption{\label{tab:Parameters} The parameters utilized in our study, extracted from references \cite{Kumar2015, Iyikanat2021}. Ref. \cite{Kumar2015} provided experimental measurements for required parameters with the exception of $\hat{\eta}$. Nevertheless, $\eta_{\parallel,z}$ can be indirectly deduced from the supplementary data contained within the same paper, specifically $\omega_{\text{LO},\parallel}$ and $\omega_{\text{LO},z}$, the frequencies where $k$ approaches infinity. 
}
\end{table}

In the limit when both phonon displacements and electromagnetic fields are small, nonlinear terms proportional to $\alpha$ and $\zeta$ in equations
(\ref{nonlinear_terms1}) and (\ref{nonlinear_terms2}) can be neglected. Assuming a homogeneous system, we take solutions in the plane wave form
\begin{align}
 & \mathbf{E}=\mathbf{E_{0}}e^{i\textbf{kr}-i\omega t},\quad \mathbf{Q}=\mathbf{Q_{0}}e^{i\textbf{kr}-i\omega t},
\end{align}
We set the $x, y$ axes to be in the hBN plane and the $z$ axis to be perpendicular to it. To find the polariton dispersion,  we fix the mode wavevector $\textbf{k}$, and solve equations (\ref{Q_EOM}) and (\ref{E_EOM}) to determine $\omega(\textbf{k})$.  Using cylindrical symmetry of the system, it is sufficient to consider the polariton wavevector of the form
\begin{align}
    \textbf{k}=k\left( 0, \cos\phi_\text{b}, \sin\phi_\text{b}\right)^T, \label{eq:k bulk}
\end{align}
Here angle $\phi_\text{b}$ denotes orientation of the wave vector in the $y - z$ plane. We use the "b" index to emphasize that here we are discussing eigenmodes in the bulk material. This should be contrasted to the case of light incident on a slab, which we will discuss later in the paper.  For our choice of the wave vector, we need to discuss separately two types of electric field polarization, either along the $x$ axis orperpendicular to it. We will refer to these two cases as s- and p-polarizations, respectively.  Although here these terms are not employed in the conventional context of light scattering from a planar sample, they will assume their natural meaning when we discuss reflectivity problem with a surface oriented along the hBN plane.

Detailed analysis of polariton modes is presented in the supplementary \cite{Suppl}, section \ref{SecSup:LinearDisp}. In the case of   s-polarization, the problem reduces to a single equation for the x-component of the electric field. In the case of p-polarization, we find coupled equations for the y- and z-components of the electric field. This can be understood as  optical anisotropy mixing TO and LO polaritons.

In Fig.~\ref{fig:Resonant couplings} we show the dispersion of p-polarized polaritons at $\phi_\text{b} =\pi/4$. We present only the real part of the frequency, while there is also a small negative imaginary part arising from phonon damping in equation (\ref{Q_EOM}). At $k = 0$, the middle and upper polaritons have finite frequencies with their reals parts expressed as,
 \begin{align}
     \omega_{0,\beta}
     = 
     \sqrt{\Omega_{\mathrm{TO},\beta}^2+\frac{\eta_\beta^2}{\epsilon_0 \epsilon_{\infty,\beta}}-\frac{\gamma_\beta^2}{4}}. \label{eq:omega _o,beta}
 \end{align}
Here the lower index $"0"$ indicates that these values correspond to the linear equations of motion. In the subsequent section, we will extend our analysis to incorporate non-linear effects, which will shift these values depending on the amplitude of the pump.

\section{Photoexcited state as Floquet Medium}
\label{sec:FloquetMedia}

In this section we generalize our discussion of hBN polaritons to the pump-probe experimental protocols, in which the pump pulse creates a large population of middle polaritons at $k=0$. We assume that pump induced state can be described using classical fields, i.e. photoexcited polaritons are in a coherent state. To understand response of the system to a weak probe pulse, we derive equations of motion for small deviations from the pump induced transient state. We make another simplifying assumption and neglect decay of pump induced polaritons during the probe pulse. Within this approximation, the pump pulse creates a Floquet medium, in which coherent oscillations of pump induced middle polaritons produce time-periodic modulation of the index of refraction of hBN. Note that under spatial inversion and reflection symmetry, the index of refraction is even, while electric field and phonon displacements are odd. Hence, on symmetry grounds index of refraction should oscillate at twice the frequency of photoexcited polaritons. The same result follows from direct consideration of nonlinear terms in equations (\ref{nonlinear_terms1}) and (\ref{nonlinear_terms2}). 

\subsection{Theoretical Framework}
\label{Sec:Pump and Probe}

We describe dynamics of the system following the pump pulse by linearizing equations of motion around the 
strong classical fields of the photoexcited middle polaritons:
\begin{eqnarray}
    &\mathbf{E} = \mathbf{E}_\text{p}e^{-i \omega_\text{p} t} + \mathbf{\delta E} +\text{c.c.}, \label{Ep+dE}\\
    &\mathbf{Q} = \mathbf{Q}_\text{p}e^{-i \omega_\text{p} t} + \mathbf{\delta Q} +\text{c.c.} \label{Qp+dQ}.
\end{eqnarray}
Here $\textbf{E}_\text{p}=E_p \hat{z}$ and $\textbf{Q}_\text{p}=Q_p \hat{z}$ represent the pump induced fields, oriented perpendicular to the hBN plane. 
$\delta \textbf{E}$ and $\delta \textbf{Q}$ represent additional fields, for example, induced by the weak probe pulse. Note that oscillation frequency of the pump field, $\omega_\text{p}$, is renormalized relative to equation (\ref{eq:omega _o,beta}) due to the large amplitude of the pump pulse (see discussion below).

We now discuss the mathematical procedure for setting up Floquet equations for  $\delta \textbf{Q}$ and $\delta \textbf{E}$. Details of the derivation are presented in in Sec. \ref{SecSup:Main_Resonances} in the supplement, and here we comment on conceptual points only. When linearizing equations (\ref{Q_EOM}) and (\ref{E_EOM}) around the time dependent solution (\ref{Ep+dE}) and (\ref{Qp+dQ}), different frequency components get mixed. We need to provide physical justification for truncating an infinite frequency ladder, which appears for linearized equations. The most important consequence of frequency mixing is parametric driving of polaritons. Mathematically, this appears as "generation" of polariton pairs due to time modulation, and physically corresponds to resonant scattering of excitations deposited into the middle polariton branch by the pump pulse. These processes give rise to parametric instabilities and amplification that we discuss below. Lowest order (and most efficient) parametric excitation of the system corresponds to creation of polariton pairs whose frequencies add up to the drive frequency $2 \omega_p$. Thus, we limit our analysis to only two frequency components of Floquet solutions, $e^{-i\omega_st}$ and $e^{i\omega_it}$, which satisfy $\omega_\text{i} + \omega_\text{s} = 2 \omega_\text{p} $. We refer to these frequency components as signal and idler components respectively. Our approximation does not include frequency mixing of the type $e^{-i\omega_st}$ and $e^{-i(\omega_s+2n\omega_p)t}$ with integer $n$. Physically, the latter type of frequency mixing corresponds to re-scattering of probe photons on the pump photons and does not lead to parametric instability or amplification.  Following these arguments, we take
\begin{align}
    &\delta \textbf{E} = \left(\textbf{E}_{\text{s}}e^{-i\omega_\text{s} t}+\textbf{E}_\text{i}^{*}e^{i\omega_\text{i}t}\right)e^{i\textbf{k}\textbf{r}}+\text{c.c.},\label{eq:delta E}\\
    &\delta \textbf{Q} = \left(\textbf{Q}_{\text{s}}e^{-i\omega_\text{s} t}+\textbf{Q}_\text{i}^{*}e^{i\omega_\text{i}t}\right)e^{i\textbf{k}\textbf{r}}+\text{c.c.} \label{eq:delta Q}.
\end{align}
Additional considerations of Floquet modes, including discussion of higher order instabilities are given in the supplement \cite{Suppl}, Sec. \ref{SecSup:Additional_Resonances}. 
 
\subsection{Nonlinear renormalization of photoexcited polaritons}

In this paper we focus on the case of the pump pulse exciting middle polaritons at $\textbf{k} =0$ polarized along the $z$-axis. This geometry retains rotational symmetry of the system in the $xy$ plane and simplifies analytical calculations. Realizing this case exactly in experiments requires sending the pump pulse from the side of hBN, which may not be practical. It is possible to excite these polaritons even when pumping is done from above the hBN plane provided that incident light comes at finite angle and is p-polarized (finite incidence angle gives nonzero in-plane momentum to polaritons, but this is not expected to alter results qualitatively). We also note that a potentially interesting approach to efficient driving of $z$-polarized polaritons is to use near-field optical systems, such as SNOM devices \cite{Ni21,Pons19,Herzig24}. 

When modeling  dynamics of polaritons excited by the pump pulse, one needs to take into account that their frequency changes as their amplitude increases. This can be described using the standard equations for a coherently driven non-linear oscillator \cite{NonlinearOscillator}. We do not present this standard analysis in this paper. However, we take into consideration the fact that frequency of photoexcited middle polaritons will depend on the amplitude of excited fields according to  (see Sec. \ref{SecSup:Pump} of the supplementary material for details). 
\begin{align}
    \omega_\text{p} \approx \omega_{0,z} \left(1 + 6\frac{ \alpha \nu_0-\zeta}{(\nu_0 \,\omega_{0,z})^2 }E_\text{p}^{2}\right) \label{eq: omega_p}
\end{align}
Here $\nu_0=\eta_{z}/ (\epsilon_{0}\epsilon_{\infty, z})$ and $\omega_{0,z}$ has been defined in Eq.\eqref{eq:omega _o,beta} with $\beta=z$. 
To provide a scale of frequency renormalization, we note that when the pump induced electric field inside hBN reaches $E_\text{p} = 1$ GV/m, which is attainable using modern technologies, the middle polariton frequency should change from $\omega_{0,z}\approx 24.9$ THz to $\omega_\text{p}\approx 25.3$ THz. This effect allows to control frequency of Floquet modulation by changing intensity of the pump pulse. 
We note that photoexcited polaritons will also have a finite decay rate arising from phonon damping. In principle, one can represent Floquet driving by a decaying mode as a multi-frequency Floquet drive. To keep discussion in this paper focused, we will assume that Floquet driving can be approximated as mono-tonal. Generalization to multi-tonal drive will be discussed elsewhere.

\subsection{Floquet eigenmodes in the bulk}

We now provide a brief overview of Floquet eigenmodes in bulk hBN for the pump pulse introduced in the previous sections. Additional details of the calculation  can be found in section \ref{SecSup:LinExpand} of the supplementary material.

Since the pump field is at zero momentum and has polarization along the $z$ axis, cylindrical symmetry is preserved. Because of this symmetry, it is sufficient to consider a Floquet eigenmode with the wave vector $\textbf{k}$ in the $yz$ plane. As in the case of equilibrium polaritons (see Section \ref{Sec:Hyperb Phon-Pol}), Floquet eigenmodes separate into two categories: 1) the signal and idler electric fields are polarized along the $x$-axis; 2) the signal and idler electric fields are polarized  perpendicular to the $x$ axis. In accordance with the discussion of equilibrium polaritons, we will refer to these cases as s- and p-polarizations, respectively.

\begin{figure*}[!t]
\begin{centering}
\includegraphics[width=0.99\textwidth]{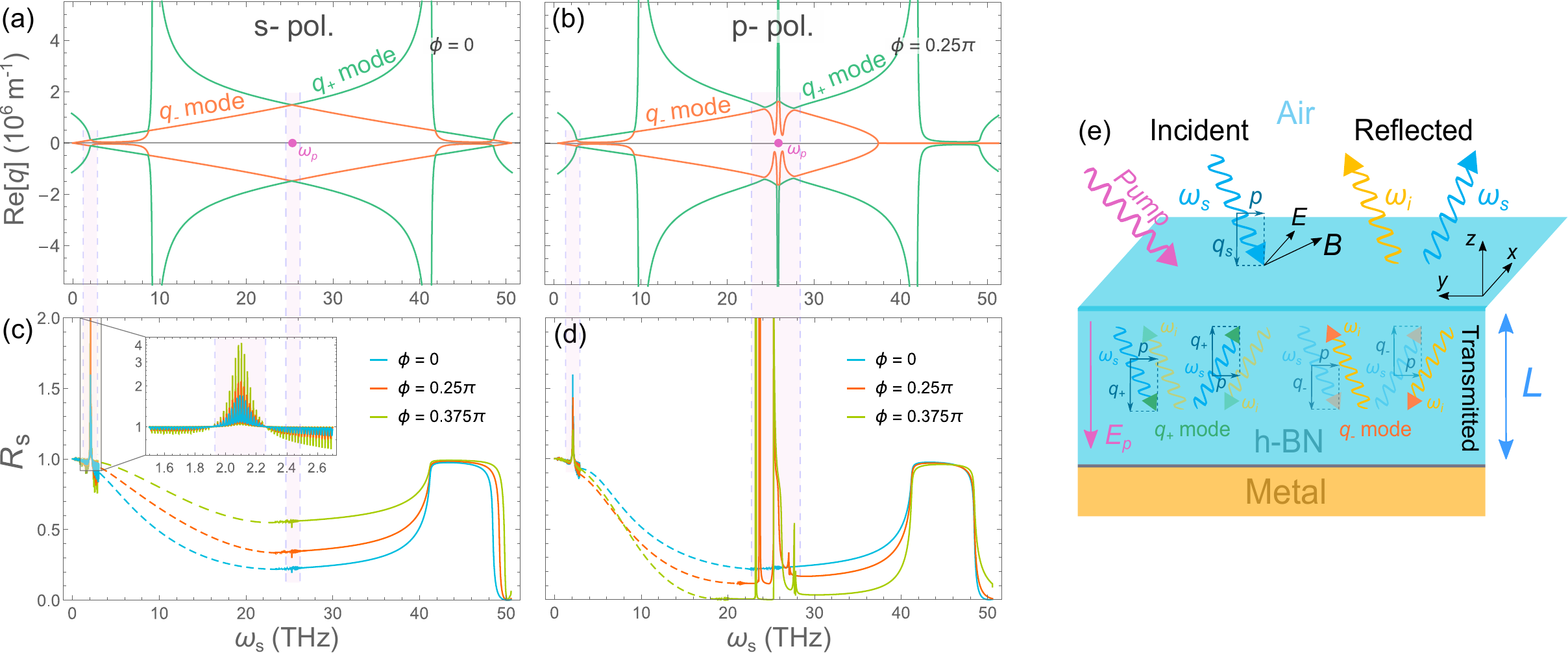}
\par\end{centering}
\caption{ \label{fig:Disp Refl} 
 (a), (b) Wave vectors of the Floquet eigenmodes in the slab as a function of the signal frequency $\omega_\text{s}$ in the presence of a pump field $E_\text{p}=1$GV/m for (a) s-polarized and (b) p-polarized incident lights. Green lines correspond to the $\pm q_+$ mode, orange to the $\pm q_-$ mode, and vertical pink strips emphasize resonant regions. In (a), electric and phonon fields lie in the $xy$ plane only, leading to two resonant scattering processes: between lower and upper polaritons (around $2, 48.5$ THz) and around the pumping frequency. In (b), for a finite angle, the $z$-polariton becomes activated, giving rise to a middle polariton allowing additional resonant scattering between middle and lower branches around $24$ and $27$ THz. (c), (d) Reflectivity versus $\omega_\text{s}$ for s- and p- polarized incident lights for a slab of thickness $L= 3$mm. Vertical pink strips connect resonant regions in dispersion curves to reflectivity plots. As can be seen in the plot at $2$ THz reflectivity significantly surpasses 100$\%$ signalling amplification. In (d), besides the $2$ THz instability, mixing between middle and upper polaritons and instability around $\omega_\text{p}$  produce reflectivity greater than 1. The oscillatory behavior observed in the inset results from constructive and destructive interference in the finite slab akin to a Fabry-P\'erot interferometer. To shift the focus of the figures to the resonances we replaced the oscillatory sections of the off-resonant regions with dashed lines. (e) Schematic illustration of the Fresnel-Floquet reflection problem. The incident signal wave excites four Floquet eigenmodes ($\pm q_{\pm}$) within the pumped hBN on top of a metal. Each mode comprises components of signal and idler frequencies. Portions of these modes are emitted back into the air, generating reflected signal and idler modes. The idler mode travels in the opposite direction to the incident signal wave.
}
\end{figure*}

We can schematically represent Floquet equations of motion as
\begin{align}
    \begin{pmatrix}\hat{D}(\omega_\text{s},\textbf{k},E_\text{p}) & E_\text{p}^2 \,\hat{M}(\omega_\text{s},\textbf{k})\\
E_\text{p}^2 \,\hat{M}(-\omega_\text{i},\textbf{k}) & \hat{D}(-\omega_\text{i},\textbf{k},E_\text{p})
\end{pmatrix}\begin{pmatrix}\textbf{E}_\text{s}\\
\textbf{Q}_\text{s}\\
\textbf{E}_\text{i}^{*}\\
\textbf{Q}_\text{i}^{*}
\end{pmatrix}=0 \label{eq:Floquet medium}
\end{align}
Matrices $\hat{D}$ and $\hat{M}$ are shown explicitly in Sec. \ref{SecSup:Floquet} of the supplementary material. In the case of s- and p-polarized modes, they are given by $2\times2$ and $4\times4$ matrices respectively. Matrix $\hat{D}$ describes light-phonon couplings for the same frequency components in the presence of the pump field, and matrix $\hat{M}$ describes mixing of signal and idler modes. Eigenvalues and eigenvectors of equation (\ref{eq:Floquet medium}) define Floquet eigenmodes of the system.

\subsection{Floquet analysis for the Fresnel reflection problem}

Even in the static case of light reflection from a planar interface (the so-called Fresnel problem), one needs to consider a problem that is reciprocal to finding polariton eigenmodes. In the latter case, one starts with a real wavevector and computes frequencies of eigenmodes. In the former case, frequency is set by incident light and one needs to determine corresponding wavevectors (generally there may be several solutions). In the Fresnel problem, wavevectors can become imaginary. This corresponds to evanescent wave solutions and occurs when incident light is in the bandgap frequency range. 

In this section we provide Floquet analysis that will be used in the next section to solve Fresnel-type reflection problems (see Ref. \cite{Marios2022} for additional discussion of this approach). We find wavevectors of Floquet eignmodes for a given frequency $\omega$. Following equation (\ref{eq:Floquet medium}), we need to look for solutions for a pair of frequency components: $\omega$ and $2\omega_p - \omega$, and two frequency components should be understood as a signal-idler pair. Said differently, a probe pulse at frequency $\omega$ triggers a response in the Floquet system at both frequencies: $\omega$ and $2\omega_p - \omega$.
Motivated by the experimental geometry in which the probe pulse is incident from above the hBN plane (see Fig. \ref{fig:Disp Refl}(e)), we assume in-plane components of momenta to be real, while $z$ components can be imaginary. This corresponds to excitation transmitted through the air-hBN interface having the same in-plane (i.e. $xy$) component of momentum as incident photons in the air. 

We take Floquet eigenmodes in the form 
\begin{align}
    \textbf{k}=\left(0,  \, p, \, q \right)^{T}, \label{eq:k refl}
\end{align}
Here, $p$ is real and will be fixed by the incident pulse (for the experimental geometry in Figure \ref {fig:Disp Refl}, $p=\frac{\omega_\text{s}}{c}\sin\phi$, where $\phi$ is the angle of incidence).
The  $z$ components of the wave vector, $q$,  should be determined by solving  Eq. \eqref{eq:Floquet medium}. Because of the signal-idler frequency mixing, for a given frequency $\omega$, we find two branches of solutions $\pm q_\pm$, as shown in Fig. \ref{fig:Disp Refl}(a),(b). Vertical pink strips highlight resonant regions where signal-idler frequency mixing is particularly efficient. Resonant regions around $2$ THz, resulting from the Floquet hybridization of the lower and upper polaritons, are evident for both s- and p- polarizations. Solutions with p- polarization show other resonance regions arising from coupling of the middle and lower branches.

To understand the nature of solutions shown in Fig. \ref{fig:Disp Refl}(a),(b), it is useful to consider first the limit of the pump amplitude going to zero. In this case the Floquet mixing matrix $\hat{M}$ vanishes, because it is proportional to $E_\text{p}^2$. Eigenmodes are determined entirely by $\hat{D}$ matrices and they decouple into two sets of solutions, corresponding to equilibrium polaritons (see Sec. \ref{Sec:Hyperb Phon-Pol}). We point out that in this case the idler wave vector can be related to the signal wave vector by reflecting around the frequency $\omega_\text{p}$. This symmetry is built into the definition of signal and idler modes, because $\omega_\text{i} = 2 \omega_\text{p} - \omega_\text{s}$.

As we introduce finite pumping strength and include Floquet mixing, elements of the  $D$ matrix have small corrections but generally remain much larger than the mixing terms in $\hat{M}$. It is possible to treat signal-idler mixing perturbatively, as long as frequencies are outside of special resonant regions. Resonant regions appear when momentum eigenvalues corresponding to  $\hat{D}(\omega_\text{s},\textbf{k},E_\text{p})$ and $\hat{D}(\omega_\text{i},\textbf{k},E_\text{p})$ become nearly equal to each other. Physically, this corresponds to momenta of the signal and idler components becoming equal (phase matching condition), which results in dramatic enhancement of Floquet parametric instability. From figures \ref{fig:Disp Refl}a, b we observe that away from the resonance regions, solutions $q_\pm$ are similar to equilibrium polaritons (note that labeling of $q_\pm$ branches in this figure does not match equilibrium polaritons, see supplementary sections for details).  On the other hand, in resonant regions, deviations can be significant.

\section{Reflectivity of Floquet system \label{Sec:Enhanced Reflectivity}}

We consider reflectivity of the system shown in figure \ref{fig:Disp Refl}. It consists of a slab of hBN
of thickness $L$ on top of a metallic cladding that we treat as a perfect mirror. We assume that the pump pulse has excited middle polariton at $q=0$ (their polarization is along the $z$ axis), so parametric driving is spatially uniform inside the slab. We consider the probe pulse incident on the
sample at an angle $\phi$ to the normal to hBN planes.

\subsection{Fresnel-Floquet reflection formalism \label{Sec:Fresnel-FLoquet}}

 In this section we show explicitly how to set up the Floquet-Fresnel problem in the case of s-polarization, when electric field is along the $x$. The case of $p$-polarization is similar, but requires keeping track of both $y$ and $z$ components of the electric field.

Above the sample, electric field is a combination of the incident and reflected fields
\begin{eqnarray}
{\bf E}_\text{air}(r,t) &=& {\bf E}_\text{inc} (r,t) + {\bf E}_\text{ref} (r,t)
\\
{\bf E}_{\rm inc} (r,t) &=& \hat{x} \, E_0 e^{ipy} e^{i q_\text{air,s} z}\, e^{ - i \omega t}  + c.c. 
\\
{\bf E}_{\rm ref} (r,t) &=& \hat{x} \, r_{\rm s} E_0 e^{ipy} e^{-i q_\text{air,s} z} \, e^{ - i \omega t}
\\
&+&\hat{x} \, r_\text{i} E_0 e^{ipy} e^{-i q_\text{air,i} z} \, e^{i (2\omega_{\rm p} - \omega) t}  + c.c. 
\end{eqnarray}
Here 
$q_{\text{air,s}}=\sqrt{\left(\omega_{\mathrm{s}}/c\right)^{2}-p^{2}}=\omega \cos \phi/c$ and $q_{\text{air,i}}=\sqrt{\left(\omega_{\mathrm{i}}/c\right)^{2}-p^{2}}$

Inside hBN electric field includes the part transmitted through the air-hBN interaface and the part reflected on the metallic mirror 
\begin{eqnarray}
{\bf E}_\text{hBN}(r,t) &=& 
\hat{x}  \lambda_{1} E_0 e^{ipy} \, e^{-i q_{+} z}\, ( \, \alpha_+  e^{ - i \omega t} + \beta_+\,e^{  i (2\omega_p - \omega) t} \,)
\nonumber\\
&+& \hat{x}  \lambda_{2} E_0 e^{ipy} \, e^{-i q_{-} z}\, ( \, \alpha_-  e^{ - i \omega t} + \beta_-\,e^{  i (2\omega_p - \omega) t} \,)
\nonumber\\
&+& 
\hat{x}  \lambda_{3} E_0 e^{ipy} \, e^{i q_{+} z}\, ( \, \alpha_+  e^{ - i \omega t} + \beta_+\,e^{  i (2\omega_p - \omega) t} \,)
\nonumber\\
&+& \hat{x}  \lambda_{4} E_0 e^{ipy} \, e^{i q_{-} z}\, ( \, \alpha_-  e^{ - i \omega t} + \beta_-\,e^{  i (2\omega_p - \omega) t} \,)
\nonumber\\
&+& c.c.
\end{eqnarray}
Here $q_{\pm}$ is obtained by solving equations \eqref{eq:Floquet medium}, as discussed in the previous section. Coefficients $\alpha_\pm$ and $\beta_\pm$ are determined by the eigenvectors of the same solution. Coefficients $\lambda_i$, $r_s$, and $r_i$ are found by imposing boundary conditions at the air/hBN and hBN/metal interfaces. Details of the calculations are presented in section \ref{SecSup:Fresnel} of the supplementary material. 

The key aspect of the Floquet-Fresnel formalism is that  Floquet eigenmodes inside hBN must be written as a combination of $\omega$ and $2\omega_\text{p}-\omega$ frequency components. As a result, reflected light in air also has both frequency components. 

\begin{figure*}[!t]
\begin{centering}
\includegraphics[width=0.99\textwidth]{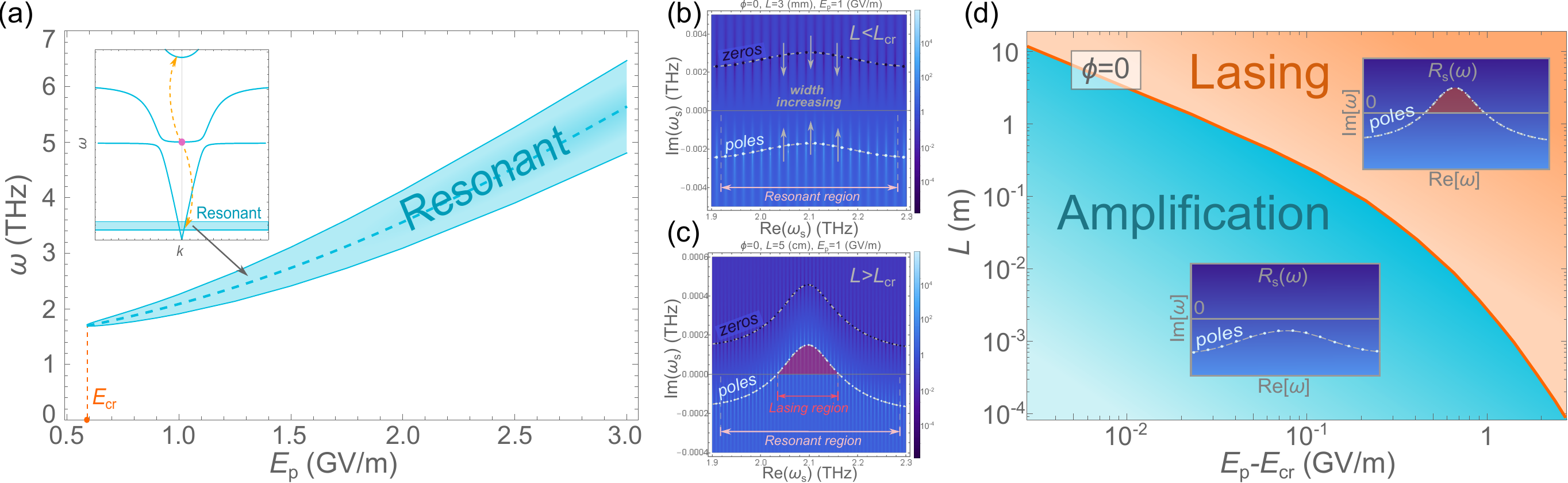}
\par\end{centering}
\caption{\label{fig:Complex_Phase_Diagram} 
Characteristics of the 2 THz resonance region exhibiting reflectivity values exceeding unity. (a) The width of the 2 THz resonant region in relation to pumping amplitude in case of a normal incidence of light $\phi=0$. The resonant region emerges when the pumping field exceeds the threshold value $E_\text{cr}\approx 0.6$ GV/m. For a stronger pump the resonant region shifts to higher frequencies and broadens, offering precise control over the resonance by modulating the pumping amplitude. The inset schematically depicts the scattering processes to the lower and upper polaritons, which give rise to this resonant region. The resonant region in the inset is extended to arbitrary wave vectors purely for illustrative purposes.
(b), (c) Depict the pole structure of reflectivity (in logarithmic scale) around $\omega_\text{s}=2.1$ THz as the width of hBN increases. (a) For a subcritical width $L=3$ mm, the poles of reflectivity are slightly shifted toward the real axis, resulting in enhanced reflectivity as seen in Fig. \ref{fig:Disp Refl}. (c) For a supercritical thickness $L=5$ cm, the poles expand into the upper half-plane, leading to lasing instabilities. In such cases and larger samples, even an infinitesimal probe pulse (or quantum fluctuations) will experience exponential amplification over time.
(d) Phase diagram: transition from amplification to lasing upon variation in slab thickness and pumping amplitude. We concentrate on the pumping fields surpassing the critical value $E_\text{cr}\approx 0.6$ GV/m. In the amplification phase, the poles reside in the lower half-plane, ensuring the system obeys causality. Increasing either the pumping or hBN thickness causes the system to cross the orange line and transition to the lasing phase, where poles migrate to the upper half-plane.
}
\end{figure*}

When analyzing the light reflection problem, we need to discuss separately cases of the system with and without parametric (lasing) instability. The onset of parametric instability is signaled by poles of the reflection coefficients $r_\text{s}$ and $r_\text{i}$ for complex $\omega$ moving into the upper half plane (see figure \ref{fig:Complex_Phase_Diagram}). To demonstrate that the same procedure can not be used in the two cases, we will discuss the question of causality of reflected light.

\subsection{Reflectivity in the absence of parametric instability}

So far we considered reflectivity coefficients for a single frequency of the probe beam. In real experiments, an optical pulse has a finite duration, which requires a finite width of the Fourier spectrum.  In the case of a system without instabilities, we decompose an incident pulse into a superposition of frequency components using the usual Fourier transformation:
\begin{align}
     &\tilde{\textbf{E}}(\textbf{r},\omega) = \int \textbf{E}(\textbf{r},t) e^{i \omega t} dt
 \end{align}
Here, frequency $\omega$ takes an arbitrary value on the real axis. For individual Fourier components, we utilize frequency-dependent reflectivity coefficients that we discussed previously. Thus for the pulse as a whole, we can write the reflected wave as:
\begin{align}
     &\textbf{E}_\text{r}(\textbf{r}, t)\nonumber\\
     &\quad=\int\tilde{\textbf{E}}(\textbf{r}, \omega)\left[r_{\mathrm{s}}(\omega) e^{-i \omega t}+r_{\mathrm{i}}(\omega) e^{i\left(2 \omega_{\mathrm{p}}-\omega\right) t}\right] \frac{d \omega}{2\pi} + c.c. \label{eq:E_r (r,t)}
\end{align}
In the last equation, integration over frequency $\omega$ goes over the entire real axis.
If we choose incident pulse such that for $t<0$ incident field is zero, then $\tilde{\textbf{E}}(\textbf{r}, \omega)$ is analytic in the upper half plane. To verify causality, one needs to demonstrate that before the incident pulse arrives, i.e. for $t<0$, reflected light is also zero.
In the stable regime, functions $r_{\mathrm{s}}(\omega)$ and $r_{\mathrm{i}}(\omega)$ are analytic in the upper half-plane. For $t<0$ we close the integration contour in (\ref{eq:E_r (r,t)}) in the upper half plane, and find that the integral vanishes because the entire integrand is analytic in the upper half plane. 

Interestingly, we find that in the stable regime but close to parametric instability, reflection coefficients can become bigger than one (see figure \ref{fig:Disp Refl} and discussion below). This indicates that our system becomes a gain medium for specific frequencies. The physical mechanism of amplification is stimulated resonant scattering. An incident probe photon stimulates production of a signal-idler pair of polaritons from the parametric driving described by Floquet equations (\ref{eq:Floquet medium}). Said differently, a probe photon stimulates a pair scattering process of the pump induced middle polaritons, in which one additional polariton at the frequency of the probe photon is produced, plus a complementary idler polariton.

\subsection{Breakdown of the naive reflectivity approach in the regime of parametric instability}

When the pump strength exceeds some critical value we find parametric lasing instability. Precise value of the drive strength depends on the thickness of the slab and is smaller for wider slabs. In the case of parametric instability, reflection coefficients have poles in the upper half plane (see figure \ref{fig:Complex_Phase_Diagram}). If we we write reflected light in the form of equation (\ref{eq:E_r (r,t)}), we will find that it violates causality. To respect causality one needs to decompose incident pulse into Fourier-like components  by doing frequency integration along a contour in the complex plane that lies above all the poles \cite{Skaar2006,Sho_22}. In the next subsection we will examine in more details how to characterize systems in the regime of instabilities.


\subsection{Effects of finite pulse duration \label{Sec:Realistic}}

To gain a deeper understanding of the system's reflectivity characteristics, we now discuss reflected light for a more realistic probe pulse shape. We assume that the probe pulse arrives at the sample at $t=0$ (i.e., $E(t<0)=0$) and has a finite duration. Our primary focus will be on the $2.1$ THz instability, so we assume that the pulse spectrum is centered around this frequency and it has a duration of approximately 250 ps. While we focus on a specific instability, the results are qualitatively applicable to all other instabilities.

\begin{figure*}[t!]
\begin{centering}
\includegraphics[width=0.99\textwidth]{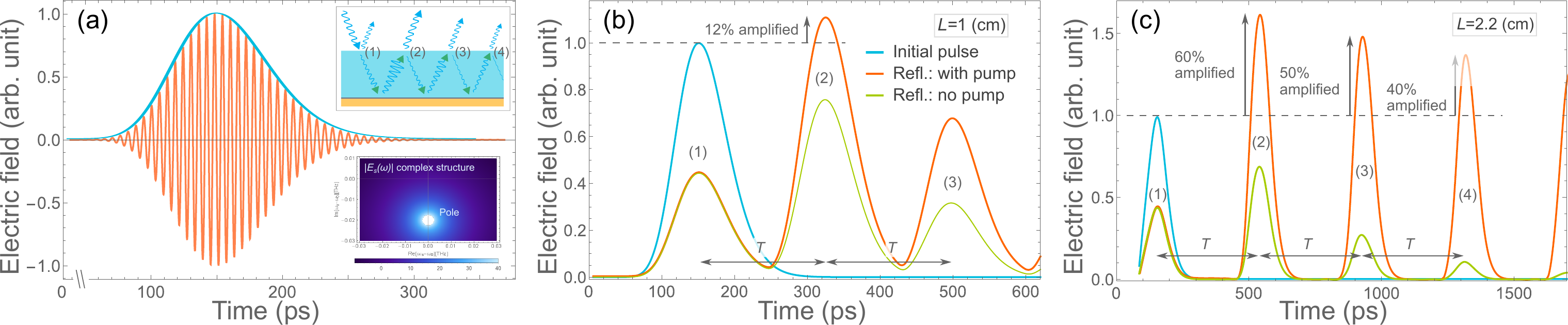}
\par\end{centering}
\caption{  (a) Electric field of a normalized realistic probe pulse operating at a frequency of $\omega_\text{s}=2.1$ THz with a duration of $0.25$ ns. The blue line traces the field amplitude, which vanishes at negative times due to causality. 
The upper inset depicts the multiple reflections experienced by the incident light. The lower inset shows the single pole of the spectral function's absolute value in the lower half-plane.
(b),(c) Amplification of a realistic pulse reflected off an hBN slab, for subcritical slab thickness values of $L=1, 2.2$ cm. 
The blue, red, and green lines represent the initial pulse amplitude, the reflected pulse with $E_\text{p}=1$ GV/m pumping, and the unpumped system, respectively. The reflected pulses exhibit multiple peaks repeating the shape of the initial pulse at specific time periods $T$.
The first peak is the initial light reflection, unaffected by pumping. The second peak occurs when light enters the hBN, reflects off the metal surface, and re-emerges after some time $T$. In the pumped case, this second peak shows notable amplification, indicating energy extraction from the pump. Later peaks gradually decay.
\label{fig:Real_pulse}}
\end{figure*}

To make sure that the incident pulse vanishes at negative times, we choose its spectral function to have poles in the lower half-plane only. For our numerical calculations, we will employ a spectral function with a single pole of order $n$, expressed as
\begin{align}
    \tilde{E}_\text{test} (\omega)= A \left(\frac{1}{\omega-\omega_\text{m} +i\Gamma}\right)^{n}, \quad \omega_\text{m}=2.1 \text{ THz}.
\end{align}
Here, the coefficient $A$ establishes the amplitude of the pulse, which we will set to unity at its maximum. The parameter $\Gamma$ governs the spectral width, and we will fix it at $\Gamma= 0.02$ THz for our calculations. Although the results should not be heavily influenced by the pole order $n$, we will ensure rapid convergence of the numerical calculations when integrals are confined to a finite region by setting the order relatively high, at $n=20$. The pulse as a function of time and its spectral properties are illustrated in Figure \ref{fig:Real_pulse}(a).

Utilizing this test probe pulse, we employ Eq. \eqref{eq:E_r (r,t)} to compute the reflected pulse, retaining only the signal term $r_\text{s}$. The numerical calculations for s-polarized normal incident light are presented in Figure \ref{fig:Real_pulse}(b),(c), with emphasis on the pulse amplitude. We analyze two subcritical slab thickness values where the system is in the amplification phase and compare the reflected pulse for pumped and unpumped systems. The calculations reveal several peaks in the reflected light, with shapes replicating the initial pulse envelope and spaced by a period dependent on the thickness. The first peak corresponds to light being reflected from the surface when the initial signal pulse strikes it. Since this pulse has not propagated substantially into the bulk of the material yet, it is unaffected by the pumping, and the first peak's pumped and unpumped reflections are identical. Yet some light passes through, propagates in the bulk, reflects off the metal sheet below, and travels back. In the amplification phase, the energy gain during this process surpasses the energy loss due to damping, resulting in the second peak exceeding $100\%$. The process continues similarly, producing multiple "aftershocks" with diminishing amplitudes, see schematic illustration in the upper inset of Fig. \ref{fig:Real_pulse}(a). As the thickness increases, the amplification becomes more pronounced, and the decay rate of the subsequent peaks weakens.

Nonetheless, the partial reflection of the initial pulse before propagating into the material's bulk wastes a significant amount of energy, necessitating a larger slab thickness to observe any amplification of the initial signal. Producing crystals of such size may be challenging. This constraint can be mitigated by slightly altering the experimental setup as discussed in \cite{Suppl}, Sec. \ref{SecSup:Cavity}.

\section{Analysis of parametric amplification and lasing}
\label{sec:Inst}

Figure \ref{fig:Disp Refl}(c),(d) presents reflectivity of the system for the same frequency as incident pulse, $R_s=|r_s|^2$, for the pump frequency $25.3$ THz. Reflectivities for both s- and p- polarizations are displayed. These results showcase that in the strongly pumped regime, reflectivity of our system can exceed unity. We find several regimes of amplification. The first instance is a pair of peaks in the reflectivity for frequencies around $24$ and $27$ THz. We observe that this is a complementary pair, in a sense of signal and idler frequencies, because the sum of the two frequencies is approximately equal to twice the pump frequency. Another singularity in reflectivity is a peak close to $2$ THz. At the complementary frequency of $48.5$ THz, singularities in the reflectivity are also present, but they are not visible in figure \ref{fig:Disp Refl}(c),(d). Finally there is an enhanced reflectivity at the frequency that is close to the pump frequency itself. Note that the $24, 27$ THz pair is exclusive to p-polarization, whereas the last two are present for both polarizations.



\subsection{2 THz Instability\label{Sec:2THz}}

The origin of reflected light amplification, and ultimately parametric instability, at $2$ THz is hybridization between the lower and upper polaritons induced by parametric driving by photoexcited middle polaritons. Microscopically, this corresponds to a resonant scattering of the pumped pair of middle polaritons into the lower and upper polariton branches shown in figure \ref{fig:Resonant couplings} with orange arrows. It is useful to note that this pair scattering process satisfies energy and momentum conservation. Reflectivity exceeds unity when the amplitude of pump induced drive exceeds the critical value $E_\text{cr}\approx 0.6$ GV/m. Finite threshold for amplification comes from the competition between parametric driving and intrinsic phonon losses.

Because of potential interest to terahertz technology \cite{THz_gap}, we present additional details of the 2 THz instability in this section. Firstly, we point out that frequency range of amplification/instability can be controlled by modifying intensity of the pumped light. Figure \ref{fig:Complex_Phase_Diagram}a) shows that as the pump field amplitude increases, the resonant region shifts to higher frequencies and concurrently expands. When amplification is present, we find it for both s- and p-polarized light, and at any finite angle of incidence. 

To understand the nature of the 2 THz instability, we analyze the pole structure of the reflectivity in the vicinity of the resonant region in figures \ref{fig:Complex_Phase_Diagram} b),c). We also use this analysis to demonstrate that  thickness of the hBN slab can be used to control parametric instability. Figure \ref{fig:Complex_Phase_Diagram} b) shows that for a slab with a thickness of $L=3$mm, poles of the reflectivity coefficient $r_\mathrm{s}(\omega)$ reside in the lower half-plane. Reflectivity also features zeros located roughly symmetrically to the poles (with respect to the real axis) in the upper half-plane. 
When reflectivity does not have poles in the upper half plane, there are no instabilities and the resonant region provides amplification without lasing. As the slab thickness increases, poles of $r_\mathrm{s}(\omega)$ start moving closer to the real axis, and above some critical thickness, $L_\text{cr}$, they cross into the upper half-plane (see Fig \ref{fig:Complex_Phase_Diagram}c)). Simultaneously, the zeros are also moving downward, and for a semi-infinite slab, they coincide with poles, forming a branch cut. Having poles in the reflectivity in the upper half plane indicates that the system is unstable to exponential growth of small fluctuations at the appropriate frequency.
This is equivalent to a lasing instability. In our case, this can be understood as the pump induced middle polaritons parametrically driving a pair of lower/upper polaritons, which results in the instability. 

Figure \ref{fig:Complex_Phase_Diagram}d) summarizes dependence of the instability regime on the pump strength and thickness of the hBN slab. It shows, that with increasing thickness of the slab, the critical value $E_\text{cr}$ deceases. We note however, that depending on geometry of the pump protocol in experiments, in thicker slabs the same power of incident pump pulse may lead to smaller amplitude of the photoexcited middle polaritons. 
Finally, we note that the resonance nature can be altered by adjusting the angle of incidence. When the probe pulse has a finite angle of incidence, it creates Floquet eigenmodes within hBN that also propagate at a finite angle. Consequently, the optical path effectively lengthens. This effect closely resembles the increase in slab thickness, leading to similar outcomes. This is discussed in more details in \cite{Suppl}, Sec. \ref{SecSup:2THz}. 

\subsection{48.5 THz Instability \label{Sec:48.5THz}}

\begin{figure}[t!]
\begin{centering}
\includegraphics[width=0.85\columnwidth]{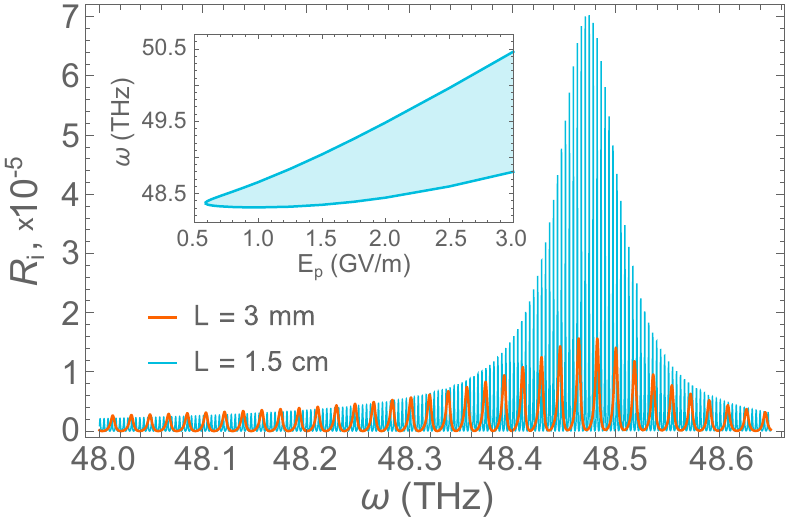}
\par\end{centering}
\caption{
The reflectivity of the idler mode $R_\text{i}$ as a function of signal frequency around the $48.5$ THz region for s-polarized incident light for two distinct slab thickness values. The prominent peak emerges due to the resonant coupling between lower and upper polaritons, as illustrated by the yellow arrows in Fig.\ref{fig:Resonant couplings}.
When probing the system at the $48.5$ THz frequency, an instability-induced idler frequency wave of $2$ THz is generated. This enables the utilization of the described system as a source of $2$ THz waves by employing a pump wave with a frequency of $\omega_\text{p}=25.3$ THz and a signal pulse at $\omega_{\text{s}}=48.5$ THz. Tuning the pumping amplitude enables control over the output idler mode frequency. The inset demonstrates the pump amplitude's influence on the region of signal frequency for resonant generation of nearly 2 THz pulse. 
The oscillatory behavior arises from constructive and destructive interference occurring within a finite slab, akin to Fabry-P\'erot interferometer.
\label{fig:Refl_49THz}}
\end{figure}

Another intriguing instability emerges at the complementary frequency to the 2 THz frequency, which is approximately $48.5$ THz. Although Figs. \ref{fig:Disp Refl}(c),(d) do not reveal any notable properties of the signal reflectivity $R_\text{s}$ in these regions, the most significant phenomenon arises when examining the reflectivity of the idler frequency rather than the signal frequency. As discussed earlier in Fig. \ref{fig:Disp Refl}(e), since the 48.5 THz signal probe pulse excites a 2 THz idler mode in hBN, some of it is emitted back into the air with reflection coefficient $r_\text{i}$.
Fig. \ref{fig:Refl_49THz} displays the idler reflectivity, $R_\text{i}=|r_\text{i}|^2$, around the anticipated unstable region, where a prominent peak is observed. By increasing the slab thickness, the reflectivity can be further enhanced. This remarkable property enables the conversion of the signal pulse into a 2 THz idler pulse.

Moreover, akin to the resonant region discussed earlier, this effect can be fine-tuned by adjusting the pump amplitude. Since this instability complements the resonant region around 2 THz depicted in Fig. \ref{fig:Complex_Phase_Diagram}, idler output at these same frequencies can be achieved using the 48.5 THz instability.  The required signal frequencies can be derived from the relation between signal and idler frequencies, $\omega_\text{s}=2\omega_\text{p}-\omega_\text{i}$, and are depicted in the inset of Fig. \ref{fig:Refl_49THz}.

Lastly, we note that this instability exhibits a pole structure similar to that observed for the 2 THz instability. 
In fact, the phase diagram illustrating the transition to lasing is identical to that for the 2 THz instability shown in Fig. \ref{fig:Complex_Phase_Diagram}(c). This is an expected result: since lasing occurs when the two polaritons from the polariton condensate generated by the pump wave break into a pair of 2 THz and 48.5 THz polaritons, if one of these frequencies features a lasing instability, the other must as well.

\subsection{Instabilities around 25 THz, and 24, 27 THz \label{sec:Other}}

\begin{figure}[t]
\begin{centering}
\includegraphics[width=0.99\columnwidth]{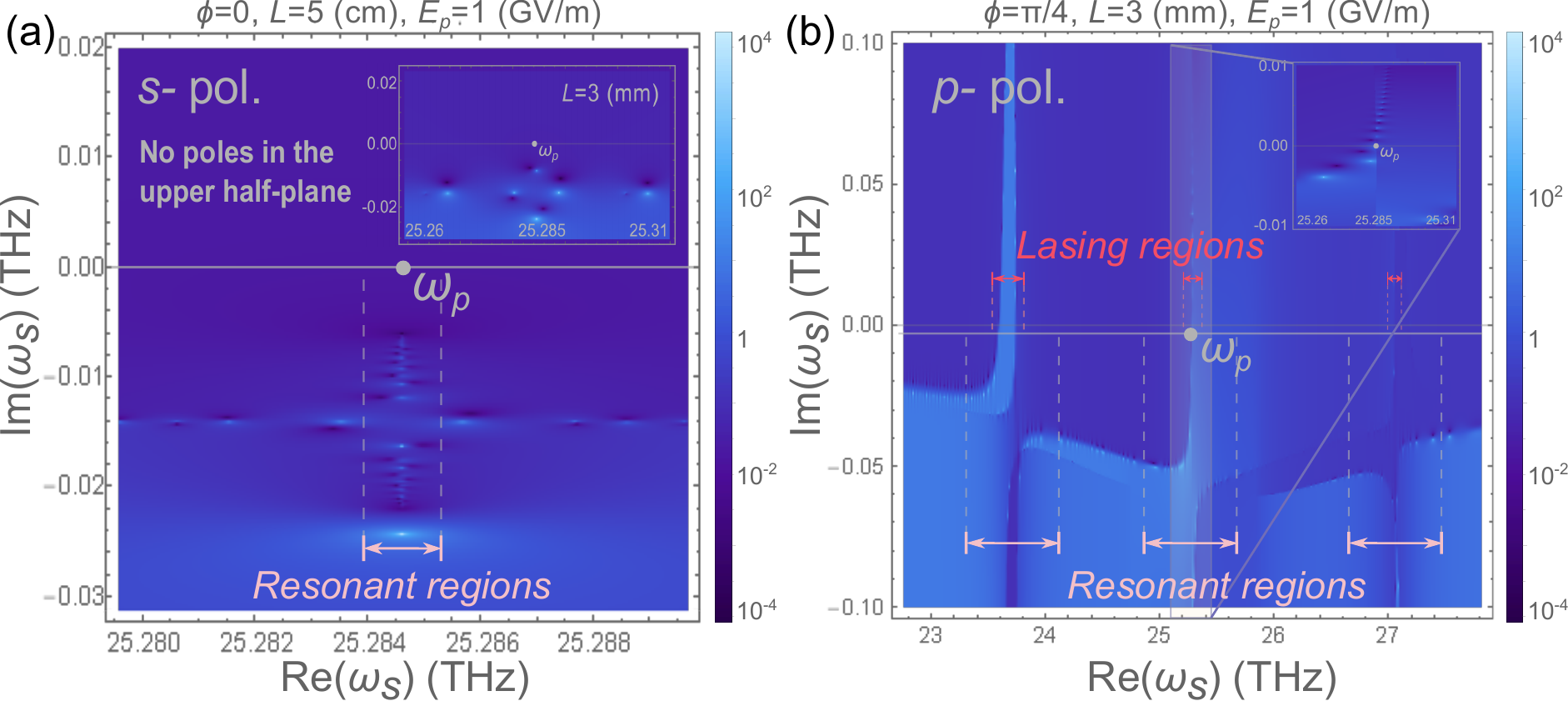}
\par\end{centering}
\caption{Poles’ structure of reflectivity (in log scale) around $\omega_\text{s} = 25$THz instability. (a) For s- polarised incident light the poles and zeros are in the lower half-plane regardless of the slab thickness. Here the resonant region does not lead to a reflectivity higher than 1 and is just denoted to the frequency region where new poles arise due to pumping. (b) For p- polarized incident light in case of a finite width the poles around 24 THz and 27 THz as well as at the pumping frequency expand to the upper half-plane creating lasing regions. \label{fig:other}}
\end{figure}

Figure \ref{fig:Disp Refl}(c),(d) demonstrates the presence of multiple resonances within the system. 
The primary instability, evident for both s- and p- polarizations across various angles, occurs near the pumping frequency $\omega_\text{p}$. These processes are depicted schematically in Figure \ref{fig:Resonant couplings} as resonant scattering processes at the middle branch with red arrows. Figure \ref{fig:Disp Refl}(c) suggests no notable phenomena in this region, aside from a minor bend. Comprehensive analysis of reflectivity in the complex plane near this point shows that the poles consistently remain in the lower half plane, regardless of the pump amplitude, slab thickness, or angle of incidence, see Fig. \ref{fig:other}(a). 

For p-polarized light, however, the situation differs significantly. In this case, poles do migrate to the upper half plane around the pumping frequency. This migration can potentially cause instability in the polariton condensate at the pump frequency, ultimately leading to its collapse. 
Moreover, for p-polarization, new instabilities emerge around $24,27$ THz as a result of lower-middle branch coupling, as illustrated in Figure \ref{fig:Resonant couplings} with green arrows. Surprisingly, these instabilities also induce lasing, even for small slab thicknesses. This is depicted in the Fig. \ref{fig:other}(b).

\section{Summary and Outlook}
\label{sec:disc}

In summary, in this paper we considered terahertz pump probe experiments in a slab of hBN material. Rich structure of phonon-polaritons in this material, resulting from its strong optical anisotropy, opens exciting prospects for observing parametric instabilities and realizing transient states with light amplification. One of the particularly encouraging findings is that a 25 THz pump can be used to achieve amplification and lasing instability for 2 THz light. 

The current work can be expanded in several directions. One crucial question that has not been addressed in our paper is developing protocols that can achieve higher pump fields. An interesting direction is to use meta-materials, such as an array of Terahertz cavities with resonant frequencies close to 25THz on the surface of hBN \cite{Berkowitz21,Mavrona21,Lee22}. These cavities will make it possible to pump with beams incident normally on hBN layers. Fringe fields of THz cavities should convert in-plane polarization of the pump pulse into z-axis polarization needed to excite middle polaritons. Furthermore, these cavities  should strengthen the amplitude of the incident pump field locally. Another interesting possibility for enhancing efficiency of the pump pulse is to use near field setups, such as SNOM devices \cite{Ni21,Pons19,Herzig24}.

The system that we discussed features several types of resonant parametric instabilities. For practical applications, the most promising is the instability involving 2 THz polaritons. We find however that this is not the dominant instability. It should be possible to extend the minimalist setup considered in this paper, e.g. using planar metamaterials \cite{Capasso2014}, in a way that strengthens the instability of the 2 THz polaritons, while suppressing the other ones. Another interesting direction is strengthening the 2THz instability by combining two or more hBN/metal pumped heterostructures. By directing an output of one into the input of the other, we can turn amplification regime into lasing but only for photons at specific frequencies.

Setup discussed in this paper can also be used to develop optical nonlinear devices at terahertz frequencies. For example, when pumping strength is close to parametric instability, reflection coefficients in some frequency range become strongly dependent on the amplitude of the pump field. Hence, addition of a small number of photons to the pump pulse can result in a bigger change in the number of reflected photons in the amplification frequency regime. This would allow to realize a terahertz optical transistor \cite{Miller2010}. Finally, we note that an essential feature of our system is generation of entangled pairs of photons. A source of entangled photons pairs in the THz regime can be an interesting addition to the toolbox of quantum communications and computations \cite{Ferguson2002, Tonouchi2007, Liebermeiste2021, Kuo2023, Kumar2022, Ahmadivand2020, Smit2019, Senica2022,Sengupta2018}.

\section{Acknowledgement}

We acknowledge useful discussions with A. Cavalleri, J. Faist, A. Imamoglu, M. Lukin, S. Yelin, P. Dolgirev and  D. Basov. K.N. acknowledges the support by  Simons Investigator Award from the Simons Foundation. I. R. acknowledges financial support in the form of a scholarship from the Den Adel Fund. 
M. H. M. acknowledges support from the Alex von Humboldt postdoctoral fellowship. E.D. acknowledges the funding by the the ARO grant number W911NF-21-1-0184, the SNSF project 200021\textunderscore212899.

\bibliography{references}

\newpage
\phantom{}
\newpage
\begin{widetext}
\renewcommand{\appendixname}{}

\appendix

	\begin{center}{\large 
	Supporting Material for ``Terahertz amplification and lasing in pump-probe experiments with hyperbolic polaritons in h-BN" }
	\end{center}
		\begin{center}{\large by Kh. Nazaryan, I. Ridkokasha, M. H. Marios and E. Demler}
	\end{center}
	\maketitle

\setcounter{figure}{0}
\renewcommand{\thefigure}{S\arabic{figure}}
\setcounter{page}{1}

\setcounter{section}{0}
\renewcommand{\thesection}{\Alph{section}}
\renewcommand{\theequation}{\thesection.\arabic{equation}}

\section{Nonlinear phonon-light interaction \label{SecSup:Hamiltonian}}
\setcounter{equation}{0}

In this section we describe the step-by-step construction of the effective Hamiltonian describing the light-phonon interaction beyond the harmonic approximation. 

HBN exhibits significant anisotropy, characterized by in-plane phonon vibrations that have a frequency nearly double that of out-of-plane phonons. We rely on ab-initio values from earlier studies \cite{Kumar2015, Iyikanat2021} to obtain precise nonlinear potentials for these modes. In the long-wavelength limit, transverse optical (TO) phonons can be represented by near-flat bands with two distinct frequencies: $\Omega_{\text{TO},\parallel}$ for the in-plane direction (along the hBN plane) and $\Omega_{\text{TO,z}}$ for the perpendicular direction. The nonlinear anisotropic Hamiltonian, which describes the TO phonons,
includes the conventional harmonic terms $H_{\text{ph}}^{(2)}$, as well as primary nonlinear terms $H_{\text{ph}}^{(4)}$, which are quartic due to the inversion symmetry exhibited by bulk hBN. Denoting $\textbf{Q}_\parallel$ for the phonon displacement parallel to the hBN plane, and $\textbf{Q}_z$ for the oscillations occurring perpendicular to the stacked hBN planes, the phononic Hamiltonian can be articulated in the following manner:
\begin{align}
&H_{ph} = H_{\text{ph}}^{(2)}+H_{\text{ph}}^{(4)}, \label{H_ph}\\
&H_{\text{ph}}^{(2)}= \frac{M \Omega_{\text{TO},\parallel}^2}{2} \mathbf{Q}_{\parallel}^2 + \frac{M \Omega_{\text{TO},z}^2}{2} \mathbf{Q}_z^2, \\
&H_{\text{ph}}^{(4)}= \zeta \mathbf{Q}^4,
\end{align}
where $M=M_B M_N /(M_B+M_N)$ is the reduced mass of the unit cell, and $\zeta$ is a parameter describing the non-linearity.  In principle, the quartic nonlinearity should also incorporate the anisotropic structure of the system, however, according to \cite{Iyikanat2021}, there is a single isotropic coefficient. Additionally, as discussed in the main text, hereafter we utilize the standard rescaling $\textbf{Q}\to \textbf{Q}/\sqrt{M}$.

Optical phonons lead to the formation of a dipole moment within the unit cell, enabling them to couple 
to light due to the interaction between the dipole moment and the electromagnetic field. Consequently, optical phonons contribute to the dielectric permittivity,  
affecting the relation between the electric field intensity and the electric induction in a monochromatic field.

In order to construct a comprehensive theory that describes the coupling between light and phonons, our complete Hamiltonian must encompass the electromagnetic energy of photons, denoted by $H_\gamma$, as well as the interaction energy, represented by $H_{\text{int}}$. Our focus lies within the terahertz (THz) frequency range, which is comparable in magnitude to phonon frequencies but is significantly smaller than characteristic electron frequencies. In other words, in this context, the electrons exhibit a static response to the field.

To account for the electromagnetic energy, we introduce $\epsilon_{\infty,\parallel}$ and $\epsilon_{\infty,z}$, which describe the dielectric responses at high frequencies in the hBN plane and in the perpendicular direction, respectively. It is important to clarify that "infinity" in this scenario refers to frequencies considerably larger than characteristic phonon frequencies but substantially smaller than characteristic electron frequencies. At these "infinite" frequencies, light interacts solely with electrons, causing the material to display dielectric responses equal to the aforementioned values. Additionally, we note that when probing with light at frequencies much higher than those of electrons (i.e., "truly infinite" frequencies), the dielectric response will be 1. 
Taking all of this into account, the electromagnetic part of the energy can be expressed as follows:
\begin{align}
    H_{\gamma}=-\frac{\epsilon_0 \left(\epsilon_{\infty,\parallel}-1\right)}{2} \textbf{E}^2_{\parallel}-\frac{\epsilon_0 \left(\epsilon_{\infty,z}-1\right)}{2} \textbf{E}^2_{z}, \label{H_gamma}
\end{align}
where $\epsilon_0$ is the permittivity of vacuum, and we introduced $\textbf{E}_{\parallel}$ and $\textbf{E}_{z}$ to denote the electric field components parallel to the hBN plane and perpendicular direction, respectively.

In addressing the interaction energy, we incorporate terms up to 4th order, consistent with the symmetry of the system
\begin{align}
    & H_{\text{int}} = H_{\text{int}}^{(2)}+ H_{\text{int}}^{(4)}, \label{H_int} 
    \quad H_{\text{int}}^{(2)}= -\eta_\parallel \textbf{Q}_\parallel \textbf{E}_\parallel -\eta_z \textbf{Q}_z \textbf{E}_z, \quad  H_{\text{int}}^{(4)} =-\alpha Q^{2} \left(\mathbf{Q}\cdot\mathbf{E}\right).
\end{align}
Here the first quadratic term describes the conventional coupling of electric field with phonon's dipole moment, which in the linear approximation is proportional to the phonon displacement vector $\textbf{Q}$ with a coefficient $\eta_\parallel$ for the in-plane component and $\eta_z$ for the out-of-plane one. The coefficient $\alpha$ signifies the non-linearity and, similarly to $\zeta$, is also isotropic.

Ultimately, the complete Hamiltonian for the system is constructed as the sum of the photonic contribution \eqref{H_ph}, the electromagnetic energy \eqref{H_gamma}, and the interaction energy \eqref{H_int}:
\begin{align}
    H= H_{\text{ph}} + H_{\gamma} + H_{\text{int}}. 
\end{align}
We summarize all the parameters used in Table \ref{tab:Parameters} of the main text.

\section{Classical equations of motion \label{SecSup:EqMot}}

Here we utilize the Hamiltonian derived in the previous section to derive the equations of motion presented in the main text. For the phononic modes we have,
\begin{align}
    &\ddot{\textbf{Q}}_\beta +\gamma_\beta \dot{\textbf{Q}}_\beta=-\frac{\partial H}{\partial \textbf{Q}_\beta} = -\tilde{\Omega}^{2}_\beta \textbf{Q}_\beta+\tilde{\eta}_\beta \textbf{E}_\beta, \label{eq:Q"_beta}\\
    &\tilde{\Omega}^{2}_\beta \left(\textbf{Q},\textbf{E}\right) = {\Omega^{2}_\beta - 4\zeta Q^{2}} +2\alpha\left(\textbf{Q}\textbf{E}\right), \\
    & \tilde{\eta}_\beta \left(\textbf{Q},\textbf{E}\right) = \eta_\beta +\alpha Q^{2}. 
\end{align}

Here, for completeness, we once again note that the coefficients $\gamma_\beta$ represent the phonons' damping coefficients, and the index $\beta = \parallel, z$ denotes the two components of the fields. As TO phonons are nearly dispersionless, we do not anticipate a significant dependence of these coefficients on phonon momentum, consequently, we will consider them as constants presented in the Table \ref{tab:Parameters}.

Despite its seemingly complex appearance, this equation has a rather straightforward interpretation. The first term in the right-hand side (RHS) implies that the phonon frequency is slightly shifted due to the nonlinear terms, a property that will play an important role in our analysis; and the second term suggests that the $\eta_\beta$ is also slightly different from the linear one. The magnitudes of these shifts are dependent on the amplitudes of the fields.

The nonlinear terms also influence the dipole moment. We compute it as follows:
\begin{align}
    \textbf{P}_\beta &=- \frac{\partial H }{\partial \textbf{E}_\beta} = \epsilon_0 \left(\epsilon_{\infty,\beta}-1\right)\textbf{E}_\beta+\tilde{\eta}_\beta \textbf{Q}_\beta
\end{align}
Subsequently, the dipole moment is incorporated into Maxwell's equations, which determine the electromagnetic field's equations of motion:
\begin{align}
\nabla \times \mathbf{E}+\partial_t \mathbf{B}=0, \quad \frac{1}{\mu_0} \nabla \times \mathbf{B} = \partial_t\left(\epsilon_0 \mathbf{E}+\mathbf{P}\right),
\end{align}
where $\mu_0$ denotes the vacuum permeability. By taking the curl of the first equation and the time derivative of the second one, and then eliminating the magnetic field, they can be combined into a single equation:
\begin{align}
\nabla_\beta \left(\nabla  \textbf{E} \right)-\nabla ^2 \textbf{E}_\beta +\frac{1}{c^2}\partial_t ^2 \left( \epsilon_{\infty,\beta} \textbf{E}_\beta +\tilde{\eta}_\beta \textbf{Q}_\beta\right) \label{eq:nabla nabla E}.
\end{align}
Consequently, Eqs. \eqref{eq:Q"_beta} and \eqref{eq:nabla nabla E} capture the nonlinear dynamics of the system.

\section{Linear Dispersions \label{SecSup:LinearDisp}}

\setcounter{equation}{0}

In this section we will elaborate the derivation of polariton spectrum in the linear regime. To that end we employ  the equations  \eqref{eq:Q"_beta} and \eqref{eq:nabla nabla E}, and neglect the non-linear terms containing $\alpha$ and $\zeta$ to  describe the linear coupling
of an electromagnetic wave propagating in the bulk material with phonons.
We look for the solution of these equations in the form of plane waves:
\begin{align} 
 & \mathbf{E}=\mathbf{E_{0}}e^{i\textbf{kr}-i\omega t},\quad\mathbf{Q}=\textbf{Q}_0 e^{i\textbf{kr}-i\omega t}
\end{align}
We can exploit the $x,y$ symmetry to choose the wave vector to be
perpendicular to the $x$ axis, hence $\mathbf{k}=\left(0,k_{y},k_{z}\right)^{T}$.

Plugging this form of solution into the equations, we can express the phonon mode through electric field by solving the equation \eqref{eq:Q"_beta}: 
\begin{align} 
 & \textbf{Q}_{0}=\frac{\hat{\eta}\textbf{E}_{0}}{\hat{g}(\omega)}, \quad \hat{g}(\omega)=\hat{\Omega}_{\text{TO}}^{2}-\omega^{2}-i\omega\hat{\gamma}
\end{align}
Here for compactness we introduced diagonal tensors $\hat{\eta}, \hat{\Omega}_\text{TO}$ and $\hat{\gamma}$ which when acting on a vector along $x,y,z$ directions apply the corresponding parameters, either in-plane for $x,y$ and out of plane for $z$.

\begin{figure*}[t]
\begin{centering}
\includegraphics[width=0.98\textwidth]{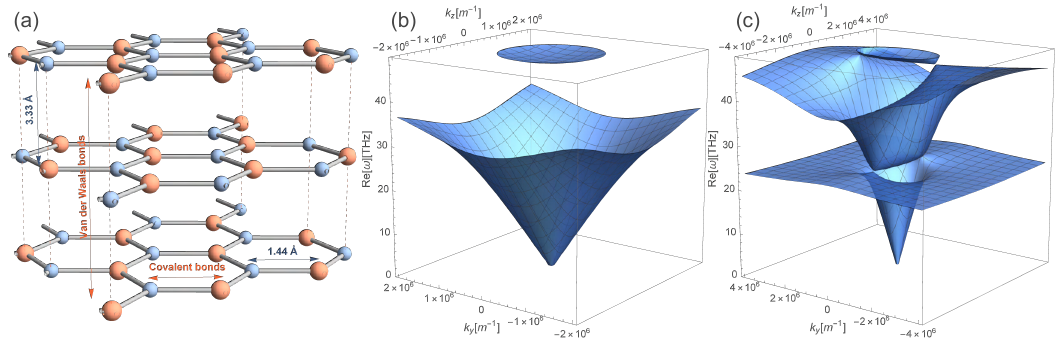}
\par\end{centering}
\caption{\label{figSup:linear} a) Schematic representation of the h-BN structure. b) Three dimensional (3D) dispersion surfaces $\text{Re}[\omega(k_{y},k_{z})]$ for the s-polarization within the range $0<\omega_\text{s}<2\omega_{0,z}\approx 50$THz, derived using Eq. \eqref{s-pol}. Two surfaces are present, each exhibiting rotational symmetry. The second surface commences just below the boundary of the considered frequency region ($\omega_{0,x}\approx 48.3$). c)  Dispersion surfaces $\text{Re}[\omega(k_{y},k_{z})]$ for the p-polarized light, as a solution to the Eq. \eqref{p-pol}. Two surfaces are present, neither of which displays rotational symmetry. The second surface starts at $\omega_{0,z}\approx24.9$ THz, while the third commences at $\omega_{0,x}$.}
\end{figure*}

We then plug this expression into the Eq. \eqref{eq:nabla nabla E} to obtain and equation governing the electric field:
\begin{align} 
 & -\left(\textbf{E}_{0},\textbf{k}\right)\textbf{k}+\textbf{E}_{0}k^{2}-\frac{\omega^{2}}{c^{2}}\left(\hat{\epsilon}_{\infty}+\frac{\hat{\eta}^{2}/\epsilon_{0}}{\hat{g}(\omega)}\right)\textbf{E}_{0}=0
\end{align}
This equation can be rewritten in a matrix form, and for a non-trivial solution for $\textbf{E}_{0}$ we require the determinant of the corresponding matrix to be zero
\begin{align} 
 & \begin{vmatrix}k^{2}-\tilde{\epsilon}_{\parallel}(\omega)\frac{\omega^{2}}{c^{2}} & 0 & 0\\
0 & k_{z}^{2}-\tilde{\epsilon}_{\parallel}(\omega)\frac{\omega^{2}}{c^{2}} & -k_{y}k_{z}\\
0 & -k_{y}k_{z} & k_{y}^{2}-\tilde{\epsilon}_{z}(\omega)\frac{\omega^{2}}{c^{2}}
\end{vmatrix}=0\label{det}
\end{align}

where we introduced $k^{2}=k_{y}^{2}+k_{z}^{2}$ and $\tilde{\epsilon}_{\alpha}(\omega)=\epsilon_{\infty,\alpha}+\frac{\eta_{\alpha}^{2}/\epsilon_{0}}{g_{\alpha}(\omega)},$
with $\alpha=\parallel,z$. 

It can be seen from \eqref{det} that thanks to the way we chose the vectors,
our problem naturally breaks down into two subproblems. The first
one is when the electric field $\textbf{E}_{0}$ is along the
$x$ axis. In this case we need only the first element in the above
matrix to be zero. The second is when $\textbf{E}_{0}$ is perpendicular
to the $x$ axis, meaning that the second block in the above matrix
should be zero. As discussed in the main text the first scenario naturally arises for an incident s-polarized wave, while the second one the p-polarized wave. Let us take advantage of this observation to rewrite the above equation as a set of two equation:
\begin{align} 
 & \text{s- polarization:} \quad k^{2}-\tilde{\epsilon}_{\parallel}(\omega)\frac{\omega^{2}}{c^{2}}=0\label{s-pol}\\
 & \text{p- polarization:}\quad \begin{vmatrix}k_{z}^{2}-\tilde{\epsilon}_{\parallel}(\omega)\frac{\omega^{2}}{c^{2}} & -k_{y}k_{z}\\
-k_{y}k_{z} & k_{y}^{2}-\tilde{\epsilon}_{z}(\omega)\frac{\omega^{2}}{c^{2}}
\end{vmatrix}=0\label{p-pol}
\end{align}


To solve these equations numerically, we use the material parameters for h-BN presented in the main text, (see Table \ref{tab:Parameters}), and present the results in the Fig. \ref{figSup:linear}. For the s- polarization we see two dispersion surfaces, both with rotational symmetry. They arise due to photon coupling solely to the phonon oscillating in-plane ($x$ direction). Meanwhile, for the p- polarization
we find three bands without rotational symmetry, indicating that photon couples to both in-plane and out-of-plane photonic modes. 
There are always 2 reststrahlen bands (gaps in dispersion
surfaces where no polaritons are available in the bulk) except the
cases when either $k_{y}=0$ or $k_{z}=0$ when one of the bands is
suppressed 
This gives a total of five polariton bands originating from the anisotropy of h-BN. This opens many interesting possibilities, for example, we
have two non-zero frequencies for $k\to 0$, meaning that we can pump
our system at two different frequencies, generating different polariton
condensates: one with pumping field along the $x$ (or $y$) axis
($\omega_{0,x}=48.3$THz), the other along the $z$ axis ($\omega_{0,z}\approx24.9$THz). The latter is the primary interest of our paper. Moreover, we note that the anisotropic properties will enable to alter significantly the reflectivity by shining probe pulse at different angles.

\section{Non-linear Theory}
\setcounter{equation}{0}

\subsection{Discussion of the main resonant coupling \label{SecSup:Main_Resonances}}

As discussed in the main text, in the pump and probe framework, we seek the solution of the non-linear equations of motion in the form
\begin{align}
    &\mathbf{E} = \mathbf{E}_\text{p}e^{-i \omega_\text{p} t} + \mathbf{\delta E} +\text{c.c.}, \label{eq:E=Ep+dE}\\
    &\mathbf{Q} = \mathbf{Q}_\text{p}e^{-i \omega_\text{p} t} + \mathbf{\delta Q} +\text{c.c.} \label{eq:Q=Qp+dQ}.
\end{align}
where $\textbf{E}_\text{p}$ and $\textbf{Q}_\text{p}$ represent the pump field and $\omega_\text{p}$ is the pump frequency. The terms $\delta \textbf{E}$ and $\delta \textbf{Q}$ are associated with the probe pulse, which is assumed to be weak, rendering these terms relatively small. Consequently, we can expand the equations of motion up to linear order with respect to these fields and formulate a linear response theory. 


We now examine the terms constituting $\delta \textbf{Q}$ (and similarly for $\delta \textbf{E}$). It should contain a wave oscillating at the signal frequency $\omega_\text{s}$, as represented by $\delta \textbf{Q}_1 \propto \left(\textbf{Q}_\text{s} e^{-i\omega_\text{s} t +i\textbf{kr}}+\text{c.c.}\right)$. To identify other frequencies present in this term, we consider a non-linear term, such as $Q^2 \textbf{Q}$ from Eq. \eqref{eq:Q"_beta}, and expand it up to linear order, resulting in terms of the following forms:
\begin{align}
    &(a) \,\, \left|Q_\text{p}\right|^2 \textbf{Q}_\text{s} e^{-i\omega_\text{s}t+i\textbf{kr}}, \quad\quad (b) \,\, \left|Q_\text{p}\right|^2 \textbf{Q}_\text{p} e^{-i\omega_\text{p}t}, \\ 
    &(c) \,\, Q_\text{p}^{*
    2} \textbf{Q}_\text{s} e^{i\left(2\omega_\text{p} - \omega_\text{s}\right)t+i\textbf{kr}}, \quad\quad (d)  \,\, Q_\text{p}^2 \textbf{Q}_\text{p} e^{-3i\omega_\text{p}t}.
\end{align}
The first two terms oscillate at signal and pump frequencies, respectively, consistent with the solution under consideration.

Conversely, term $(c)$ introduces a new active frequency,
\begin{align}
    \omega_\text{i}=2\omega_{\text{p}}-\omega_{\text{s}}, \label{eq:omega_i}
\end{align}
referred to as the idler frequency. The solution  $\delta \textbf{Q}$ must encompass this frequency as well, since the coupling of the signal frequency to idler one implies that the term with idler frequency in $\delta \textbf{Q}$ would generate additional terms at signal frequency, and vice versa. This term is the primary focus in the main text. 
We also note that in order to avoid having negative idler frequencies, we will limit our discussion to the region $0<\omega_\text{s}<2\omega_\text{p}$.

From the form of these terms, we can also deduce that the idler mode should possess a wave vector opposite to the signal one, given by 
\begin{align}
    \textbf{k}_\text{i}=-\textbf{k}_\text{s} \label{eq:q_i}
\end{align}
As a result, we observe effective energy and momentum conservation laws in this system. 

Considering term (d), which oscillates at triple pump frequency, we reference the conventional non-linear classical oscillator problem. In this context, we know that the triple frequency term does not produce secular terms and results in a weak additional mode at $3\omega_\text{p}$ frequency. This mode is not our primary focus, allowing us to exclude these terms from our discussion.

\subsection{Discussion of additional resonant couplings \label{SecSup:Additional_Resonances}}

In the main body of this paper, our analysis was primarily centered on the resonant couplings involving an idler mode, which is connected to the signal mode via relation, 
\begin{align} 
    (\text{I}) \quad \omega_\text{i}=2\omega_\text{p}-\omega_\text{s}. \label{eqSup:I omega_i}
\end{align}
As discussed, this process leads to a parametric resonance and is the primary focus of our analysis.  
In this section our objective is to discuss additional type of processes that can arise in the effective Floquet medium. 

Firstly, from the analysis in the previous section \ref{SecSup:Main_Resonances}, we can deduce the existence of other terms, given by the form $Q_\text{p}^{2} \textbf{Q}_\text{s} e^{-i\left(2\omega_\text{p} + \omega_\text{s}\right)t+i\textbf{kr}}$. In these processes, a signal and two pump polaritons combine to create a new mode. Therefore, this type of processes lead to a \textit{scattering} of the signal wave to $\omega_\text{s}^\prime = 2\omega_\text{p}+ \omega_\text{s}$, rather than producing a parametric resonance. This concludes all the terms that arise in the leading order of perturbation theory $\propto Q_\text{p}^2 \delta Q$.

To obtain additional parametric resonances we
need to consider higher order terms, $\propto Q_\text{p}^{2n} \delta Q$. They will lead to signal and idler frequencies differing by $2n\omega_{\text{p}}$ with any integer $n>1$. Yet, they are of higher order in pump field and thus are parametrically smaller.



Now we will conduct a qualitative analysis of these instabilities to demonstrate that the instabilities discussed above will not interfere with the primary instability (I). Let us choose the following two for illustration,
\begin{align} 
    (\text{II}) \quad \text{scattering:}\quad \omega_\text{s}^\prime = 2\omega_\text{p}+ \omega_\text{s}; \quad (\text{III}) \quad \text{higher order:} \quad \omega_\text{i}^{(4)}=4\omega_\text{p}-\omega_\text{s}. \label{eqSup:II III omega_i}
\end{align}

For a qualitative analysis, we can plot the signal and complimentary (scattered for (II) or idler fir (III)) mode dispersion curves concurrently. The complementary mode can be derived using the equations above to express them through the signal one. The regions where the two curves intersect correspond to the resonant regions. The resonant processes for all three types of couplings are illustrated in Fig. \ref{figSup:New_type_instabilities}. A crucial observation is that the resonances resulting from the (II) and (III) type processes are significantly distanced from those associated with the type (I) processes, particularly from the principal 2 THz (and complementary) instability. This supports the legitimacy of our approach of disregarding these other types of instabilities when analyzing the main one.

Furthermore, we note that for the (II) and (III) type processes, the resonant couplings generally occur at high momenta. In a realistic system, phononic dispersion will begin to deviate from the flat band approximation used in our work, leading to an increase in their damping coefficients with momentum as well. We posit that this effect will mitigate the impact of these additional resonances.

\begin{figure*}[h]
\begin{centering}
\includegraphics[width=0.98\textwidth]{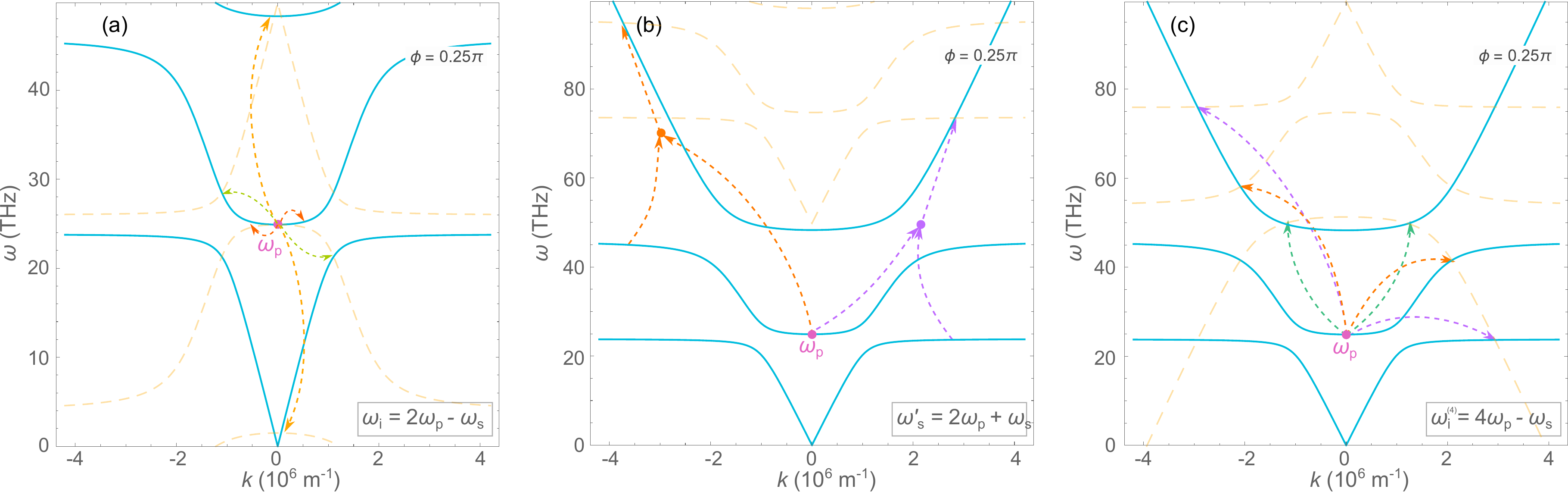}
\par\end{centering}
\caption{Schematic representations of resonant couplings corresponding to different idler modes. Blue line represents the polariton spectrum at signal frequency, while the dashed yellow line is for idler component defined through relations \eqref{eqSup:I omega_i} or \eqref{eqSup:II III omega_i} depending in the process type.    
a) Type (I) processes detailed in the main text. Here two polaritons at pump frequency break down into a pair of signal and idler frequencies.  b) In type (II) processes two polaritons from condensate get combined with signal polariton to produce an idler one. c) Type (III) processes, arise when including higher order driving modes. In this case pumped polaritons break into a pair which possess frequencies with sum of $4\omega_\text{p}$. We see that in type (II) and (III) processes the resonances typically occur at large wave vectors right at the edge of restrahlen bands \label{figSup:New_type_instabilities}}
\end{figure*}

\subsection{\label{SecSup:LinExpand} Linear expansion of non-linear terms}

In the main text we employed the solution ansatz \eqref{eq:E=Ep+dE},\eqref{eq:Q=Qp+dQ} to linearize the non-linear terms in the equations of motion. The solution consists of three modes and for convenience we rewrite it here again,
\begin{align} 
 & \textbf{Q}=\mathbf{Q}_\text{p}e^{-i\omega_\text{p}t}+c.c+\left(\mathbf{Q}_\text{s}e^{-i\omega_\text{s}t}+\mathbf{Q}_\text{i}^*e^{i\omega_\text{i}t}\right)e^{i\mathbf{k}\mathbf{r}}+c.c\\ 
& \textbf{E}=\mathbf{E}_\text{p}e^{-i\omega_\text{p}t}+c.c+\left(\mathbf{E}_\text{s}e^{-i\omega_\text{s}t}+\mathbf{E}_\text{i}^*e^{i\omega_\text{i}t}\right)e^{i\mathbf{k}\mathbf{r}}+c.c
\end{align}

In this section we present the explicit expressions of the nonlinear terms after
the substituting this solution and preserving only the terms linear
in signal and idler fields:
\begin{align} 
    Q^2\mathbf{E}=	&\left[Q_\text{p}^{2}\mathbf{E}_\text{p}^*+2|Q_\text{p}|^{2}\mathbf{E}_\text{p}\right]e^{-it\omega_\text{p}}\nonumber\\
    &+\left[Q_\text{p}^{2}\mathbf{E}_\text{i}^*+2|Q_\text{p}|^{2}\mathbf{E}_\text{s}+2\left(\mathbf{Q}_\text{p}\cdot\mathbf{Q}_\text{s}\right)\mathbf{E}_\text{p}^*+2\left(\mathbf{Q}_\text{p}^{*}\cdot\mathbf{Q}_\text{s}\right)\mathbf{E}_\text{p}+2\left(\mathbf{Q}_\text{p}\cdot \mathbf{Q}_\text{i}^{*}\right)\mathbf{E}_\text{p}\right]e^{i\mathbf{kr}-it\omega_\text{s}} \nonumber\\
    &+\left[Q_\text{p}^{*2}\mathbf{E}_\text{s}+2|Q_\text{p}|^{2}\mathbf{E}_\text{i}^*+2\left(\mathbf{Q}_\text{p}\cdot\mathbf{Q}_\text{i}^{*}\right)\mathbf{E}_\text{p}^*+2\left(\mathbf{Q}_\text{p}^{*}\cdot\mathbf{Q}_\text{i}^{*}\right)\mathbf{E}_\text{p} +2\left(\mathbf{Q}_\text{p}^{*}\cdot\mathbf{Q}_\text{s}\right)\mathbf{E}_\text{p}^*\right]e^{i\mathbf{kr}+it\omega_\text{i}}+c.c. \\ 
        Q^2\mathbf{Q}=	&\left[Q_\text{p}^{2}\mathbf{Q}_\text{p}^*+2|Q_\text{p}|^{2}\mathbf{Q}_\text{p}\right]e^{-it\omega_\text{p}}\nonumber\\
    &+\left[Q_\text{p}^{2}\mathbf{Q}_\text{i}^*+2|Q_\text{p}|^{2}\mathbf{Q}_\text{s}+2\left(\mathbf{Q}_\text{p}\cdot\mathbf{Q}_\text{s}\right)\mathbf{Q}_\text{p}^*+2\left(\mathbf{Q}_\text{p}^{*}\cdot\mathbf{Q}_\text{s}\right)\mathbf{Q}_\text{p}+2\left(\mathbf{Q}_\text{p}\cdot \mathbf{Q}_\text{i}^{*}\right)\mathbf{Q}_\text{p}\right]e^{i\mathbf{kr}-it\omega_\text{s}} \nonumber\\
    &+\left[Q_\text{p}^{*2}\mathbf{Q}_\text{s}+2|Q_\text{p}|^{2}\mathbf{Q}_\text{i}^*+2\left(\mathbf{Q}_\text{p}\cdot\mathbf{Q}_\text{i}^{*}\right)\mathbf{Q}_\text{p}^*+2\left(\mathbf{Q}_\text{p}^{*}\cdot\mathbf{Q}_\text{i}^{*}\right)\mathbf{Q}_\text{p} +2\left(\mathbf{Q}_\text{p}^{*}\cdot\mathbf{Q}_\text{s}\right)\mathbf{Q}_\text{p}^*\right]e^{i\mathbf{kr}+it\omega_\text{i}}+c.c.\\ 
	\left(\mathbf{Q}\cdot\mathbf{E}\right)\mathbf{Q}=	&\left[\left(\mathbf{Q}_\text{p}\cdot\mathbf{E}_\text{p}^{*}\right)\mathbf{Q}_\text{p}+\left(\mathbf{Q}_\text{p}\cdot\mathbf{E}_\text{p}\right)\mathbf{Q}_\text{p}^*+\left(\mathbf{Q}_\text{p}^{*}\cdot\mathbf{E}_\text{p}\right)\mathbf{Q}_\text{p}\right]e^{-it\omega_\text{p}}+\nonumber\\
	&+\big[\left(\mathbf{Q}_\text{p}\cdot\mathbf{E}_\text{i}^{*}\right)\mathbf{Q}_\text{p}+\left(\mathbf{Q}_\text{p}\cdot\mathbf{E}_\text{s}\right)\mathbf{Q}_\text{p}^*+\left(\mathbf{Q}_\text{p}^{*}\cdot\mathbf{E}_\text{s}\right)\mathbf{Q}_\text{p}+\left(\mathbf{Q}_\text{p}\cdot\mathbf{E}_\text{p}^{*}\right)\mathbf{Q}_\text{s}+\left(\mathbf{Q}_\text{s}\cdot\mathbf{E}_\text{p}^{*}\right)\mathbf{Q}_\text{p}+\nonumber\\
	&\quad\quad+\left(\mathbf{Q}_\text{p}^{*}\cdot\mathbf{E}_\text{p}\right)\mathbf{Q}_\text{s}+\left(\mathbf{Q}_\text{s}\cdot\mathbf{E}_\text{p}\right)\mathbf{Q}_\text{p}^*+\left(\mathbf{Q}_\text{p}\cdot\mathbf{E}_\text{p}\right)\mathbf{Q}_\text{i}^*+\left(\mathbf{Q}_\text{i}^{*}\cdot\mathbf{E}_\text{p}\right)\mathbf{Q}_\text{p}\big]e^{ikr-it\omega_\text{s}}+\nonumber\\
	&+\big[\left(\mathbf{Q}_\text{p}^{*}\cdot\mathbf{E}_\text{s}\right)\mathbf{Q}_\text{p}^*+\left(\mathbf{Q}_\text{p}\cdot\mathbf{E}_\text{i}^{*}\right)\mathbf{Q}_\text{p}^*+\left(\mathbf{Q}_\text{p}^{*}\cdot\mathbf{E}_\text{i}^{*}\right)\mathbf{Q}_\text{p}+\left(\mathbf{Q}_\text{p}\cdot\mathbf{E}_\text{p}^{*}\right)\mathbf{Q}_\text{i}^{*}+\left(\mathbf{Q}_\text{i}^{*}\cdot\mathbf{E}_\text{p}^{*}\right)\mathbf{Q}_\text{p}+\nonumber\\
	&\quad\quad+\left(\mathbf{Q}_\text{p}^{*}\cdot\mathbf{E}_\text{p}\right)\mathbf{Q}_\text{i}^*+\left(\mathbf{Q}_\text{i}^{*}\cdot\mathbf{E}_\text{p}\right)\mathbf{Q}_\text{p}^*+\left(\mathbf{E}_\text{p}^{*}\cdot\mathbf{Q}_\text{s}\right)\mathbf{Q}_\text{p}^*+\left(\mathbf{E}_\text{p}^{*}\cdot\mathbf{Q}_\text{p}^{*}\right)\mathbf{Q}_\text{s}\big]e^{ikr+it\omega_\text{i}}+c.c.  
\end{align}

We note that this expansion also includes only the three frequencies, $\omega_{\text{p}}$, $\omega_{\text{s}}$, $\omega_{\text{i}}=2\omega_{\text{p}}-\omega_{\text{s}}$. 
We then utilize this expansion to separate the full equation of motion into three separate ones corresponding to each of these frequencies. 

In the subsequent sections we will analyze the equations for each modes, focusing on specific cases of s- and p- polarized light.  

\subsection{Pump wave \label{SecSup:Pump}}

Here we detail the equations of motion governing the pump wave dynamics. We employ the expansion provided above and retain only the terms oscillating at the pump frequency. 

Following the discussion from the main text we orient the pump electric field along the $z$-axis. Due to the rotational symmetry, the associated phononic field points along $z$ axis as well, $E_{p}\parallel Q_{p}$. Thus we obtain the following equations	
\begin{align}  
    &g_{z}(\omega_{p})Q_{p}-\eta_{z}E_{\text{p}}=3\alpha\left(Q_{\text{p}}^{2}E_{\text{p}}^{*}+2|Q_\text{p}|^{2}E_{\text{p}}\right)+12\zeta|Q_\text{p}|^{2} Q_{\text{p}}\\
     &\epsilon_{\infty,z}E_{\text{p}}+\frac{\eta_{z}}{\epsilon_{0}}Q_{\text{p}}+\frac{3\alpha}{\epsilon_{0}} |Q_{\text{p}}|^{2}Q_{\text{p}}=0
\end{align}
Let us rewrite, assuming $Q_\text{p}=- E_\text{p}/\nu$ with a real $E_\text{p} \in\mathbb{R}$. This rewrites the second equation as follows,	
\begin{align}  
    &g_{z}(\omega_{p})+\eta_{z}\nu+\frac{3\alpha}{\nu}\left(1+2\frac{\nu^2}{|\nu|^{2}}\right)E_{\text{p}}^{2}-\frac{12\zeta}{|\nu|^{2}}E_{\text{p}}^{2}=0\\
     &\nu \epsilon_{\infty,z}-\frac{\eta_{z}}{\epsilon_{0}}-\frac{3\alpha}{\epsilon_{0}}\frac{1}{|\nu|^{2}}E_{\text{p}}^{2}=0
\end{align}
The second equation has a single real solution for $\nu$, which in zero pumping limit tends to $\nu_0= \eta_{z}/(\epsilon_{0}\epsilon_{\infty z})$. We then plug that solution into the first equation to derive the pumping frequency as a function of pumping amplitude.  The Taylor expansion of the latter in the leading order in pumping produces the result \eqref{eq: omega_p} introduced in the main text. 

\subsection{\label{SecSup:Floquet} Floquet eigenmodes}

In this section we discuss the derivations of equations describing the propagation of emerging Floquet eigenmodes in the cases of s- and p- polarized waves.

\subsection*{s- polarization}

In the case of s- polarized light we choose $\mathbf{E}_{\text{s}/\text{i}}$ along the $x$ axis and employ the linear expansion of the non-linear terms derived in the Sec.\ref{SecSup:LinExpand}. Somewhat lengthy but straightforward manipulations rewrite the equations in the form, introduced in the main text,
\begin{align}  
    \begin{pmatrix}\hat{D}^{(s)}(\omega_\text{s},\textbf{k},E_\text{p}) & E_\text{p}^2 \,\hat{M}^{(s)}(\omega_\text{s},\textbf{k})\\
E_\text{p}^2 \,\hat{M}^{(s)}(-\omega_\text{i},\textbf{k}) & \hat{D}^{(s)}(-\omega_\text{i},\textbf{k},E_\text{p})
\end{pmatrix}
 \begin{pmatrix}
    E_{sx}\\
    Q_{sx}\\
    E_{ix}^{*}\\
    Q_{ix}^{*}
\end{pmatrix}.
\end{align}
Here the matrices are expressed explicitly as
\begin{align} 
    \hat{D}^{(s)}(\omega,\textbf{k},E_\text{p})= \begin{pmatrix}k^{2}-\frac{\omega^{2}}{c^{2}}\epsilon_{\infty} & -\frac{\omega^{2}}{c^{2}\epsilon_{0}}\left(\eta_x+2\frac{\alpha}{\nu^2} E_{p}^{2}\right)\\
-\eta_x-\frac{2\alpha}{\nu^2} E_{p}^{2} & \hat{g}(\omega)+\frac{4}{\nu^2}E_{p}^{2}\left(\alpha\nu-2\zeta\right)
\end{pmatrix} ,\quad \hat{M}^{(s)}(\omega,\textbf{k})=\frac{1}{\nu^{2}}\begin{pmatrix}0 & -\frac{\omega^{2}}{c^{2}}\frac{\alpha}{\epsilon_{0}}\\
-\alpha & 2\left(\alpha\nu-2\zeta\right)
\end{pmatrix}
\end{align}
 The matrix $\hat{D}$ characterizes the coupling between light and phonon within a given mode in the presence of the pumping field. 
The matrix $\hat{M}$ accounts for the mixing between signal and idler modes. The Floquet eigenmodes are subsequently determined by equating the determinant of the matrix from the aforementioned equation to zero.  As a result we obtain two pairs solutions $\pm q_\pm (\omega_\text{s})$, which are illustrated in the Fig. \ref{fig:Disp Refl}a) in the main text. Additionally, the eigenmodes allow us to link the idler component to the signal one as $E_{\text{i}x}^\pm = \xi_{\text{i}1/2x}^\pm E_{\text{s}x}^\pm$, where the indices $1,2$ refer to the two eigenmodes in each pair $\pm q$ differing by the sign. Here the $\xi$-s are dimensionless parameters.

\subsection*{p- polarization}

The equations for the p- polarized wave are derived in a similar manner by orienting the fields in the $y-z$ plane. The resulting equations read as follows,

\begin{align} 
    \begin{pmatrix}\hat{D}^{(p)}(\omega_\text{s},\textbf{k},E_\text{p}) & E_\text{p}^2 \,\hat{M}^{(p)}(\omega_\text{s},\textbf{k})\\
E_\text{p}^2 \,\hat{M}^{(p)}(-\omega_\text{i},\textbf{k}) & \hat{D}^{(p)}(-\omega_\text{i},\textbf{k},E_\text{p})
\end{pmatrix}
 \begin{pmatrix}
    E_{sy}\\
    Q_{sy}\\
    E_{sz}\\
    Q_{sz}\\
    E_{iy}^{*}\\
    Q_{iy}^{*}\\
    E_{iz}^{*}\\
    Q_{iz}^{*}
\end{pmatrix} = 0. 
\end{align}


The matrices in this case are $4\times 4$ and are expressed as
\begin{align} 
&\hat{D}^{(p)}(\omega,\textbf{k},E_\text{p}) = \begin{pmatrix}k_{z}^{2}-\frac{\omega^{2}}{c^{2}}\epsilon_{\infty y} & -\frac{\omega^{2}}{c^{2}}\frac{\eta_{y}}{\epsilon_{0}}-2\frac{\alpha}{\nu^{2}\epsilon_{0}}\frac{\omega^{2}}{c^{2}}E_\text{p}^{2} & -k_{y}k_{z} & 0\\
\eta_{y}+2\frac{\alpha}{\nu^{2}}E_\text{p}^{2} & -g_{y}(\omega)+4\frac{2\zeta-\alpha\nu}{\nu^{2}}E_\text{p}^{2} & 0 & 0\\
-k_{y}k_{z} & 0 & k_{y}^{2}-\frac{\omega^{2}}{c^{2}}\epsilon_{\infty z} & -\frac{\omega^{2}}{c^{2}}\frac{\eta_{z}}{\epsilon_{0}}-6\frac{\alpha}{\nu^{2}\epsilon_{0}}\frac{\omega^{2}}{c^{2}}E_\text{p}^{2}\\
0 & 0 & \eta_{z}+6\frac{\alpha}{\nu^2} E_\text{p}^{2} & -g_{z}(\omega)+12\frac{2\zeta-\alpha\nu}{\nu^{2}}E_\text{p}^{2}
\end{pmatrix};\\
    &\hat{M}^{(p)}(\omega,\textbf{k}) =\frac{1}{\nu^{2}}\begin{pmatrix}0 & -\alpha\frac{\omega^{2}}{c^{2}\epsilon_0} & 0 & 0\\
\alpha & 2\left(2\zeta-\alpha\nu\right) & 0 & 0\\
0 & 0 & 0 & -3\alpha\frac{\omega^{2}}{c^{2}\epsilon_{0}}\\
0 & 0 & 3\alpha & 6\left(2\zeta-\alpha\nu\right)
\end{pmatrix}.
\end{align}

By finding the eigenvalues of the matrix in the above equation we derive the Floquet eigenstates of the driven media, depicted in the Fig. \ref{fig:Disp Refl} of the main text. 
Also, similarly to the s- polarization case, we find the eigenvectors of the corresponding eigenstates to obtain the relation between the different field components. Specifically, we express them through $E_{\text{s}y}$, such as $E_{sz}^{+}=\xi_{sz}^{+}E_{sy}^{+}$, $E_{\text{i}y}^{*+}=\xi_{\text{i}y}^{+}E_{\text{s}y}^{+}$ and
$E_{\text{i}z}^{*+}=\xi_{\text{i}z}^{+}E_{\text{s}y}^{+}$.

\section{\label{SecSup:Fresnel} Fresnel-Floquet equations}
\setcounter{equation}{0}

Here we provide the detailed elaboration of the Fresnel-Floquet reflection problem for s- and p- polarized incident lights. The analysis utilizes the Floquet eigenmodes described in the previous section and links them to the incident light.

\subsection{s- polarized light}

For the s- polarized incident light, the electric field points along the $x$ axis along the surface. It can be expressed explicitly in the air and within the slab (leaving out the time dependent terms $e^{-i\omega t}$ and focusing on the spatial dependence), 
\begin{align} 
    &\mathbf{E}_{\mathrm{s}}(\mathbf{r})=\left\{ \begin{array}{ll}
E_{0}\hat{\mathbf{x}}e^{ipy}\left(e^{-iq_{\mathrm{s}}z}+r_{\mathrm{s}}e^{iq_{\mathrm{s}}z}\right) & (0<z)\\
E_{0}\hat{\mathbf{x}}e^{ipy}\left(t_{s1+}e^{iq_{+}z}+t_{s2+}e^{-iq_{+}z}+t_{s1-}e^{iq_{-}z}+t_{s2-}e^{-iq_{-}z}\right) & (z<0)
\end{array}\right.\\
&\mathbf{E}_{\mathrm{i}}^{*}(\mathbf{r})=\left\{ \begin{array}{llc}
E_{0}\hat{\mathbf{x}}e^{ipy}\left(r_{\mathrm{i}}e^{-iq_{\mathrm{i}}z}\right) &  & (0<z)\\
E_{0}\hat{\mathbf{x}}e^{ipy}\left(t_{i1+}e^{iq_{+}z}+t_{i2+}e^{-iq_{+}z}+t_{i1-}e^{iq_{-}z}+t_{i2-}e^{-iq_{-}z}\right) &  & (z<0)
\end{array}\right.
\end{align}

Here, $q_{\mathrm{s}z}=\sqrt{\left(\omega_{\mathrm{s}}/c\right)^{2}-p^{2}},q_{\mathrm{i}z}=\sqrt{\left(\omega_{\mathrm{i}}/c\right)^{2}-p^{2}}$, and we introduced $t_{\text{s/i}1\pm},t_{\text{s/i}2\pm}$ for the transmission coefficients of the different active modes, as well as $r_{\text{s/i}}$ for the reflection coefficients of signal and idler components. 

We now employ the Maxwell's equation,
\begin{align} 
    \nabla\times\mathbf{E}+\left(-i\omega \right)\mathbf{B}=0\Rightarrow\mathbf{B}=-i\frac{\nabla\times \textbf{E} }{\omega} \label{eqSup:MaxSup}.
\end{align}
to recover the magnetic field. The latter is expressed as follows,
\begin{align} 
    &\mathbf{B}_{\mathrm{s}}(\mathbf{r})=\left\{ \begin{array}{ll}
-\frac{E_{0}}{\omega_{\mathrm{s}}}e^{ipy}\left(e^{-iq_{\mathrm{s}}z}\left(q_{\mathrm{s}}\hat{\mathbf{y}}+p\hat{\mathbf{z}}\right)+r_{\mathrm{s}}\left(-q_{\mathrm{s}}\hat{\mathbf{y}}+p\hat{\mathbf{z}}\right)e^{iq_{\mathrm{s}}z}\right) & (0<z)\\
-\frac{E_{0}}{\omega_{\mathrm{s}}}e^{ipy}\left(t_{\mathrm{s}1+}\left(-q_{+}\hat{\mathbf{y}}+p\hat{\mathbf{z}}\right)e^{iq_{+}z}+t_{\mathrm{s}2+}\left(q_{+}\hat{\mathbf{y}}+p\hat{\mathbf{z}}\right)e^{-iq_{+}z}+t_{\mathrm{s}1-}\left(-q_{-}\hat{\mathbf{y}}+p\hat{\mathbf{z}}\right)e^{iq_{-}z}+t_{\mathrm{s}2-}\left(q_{-}\hat{\mathbf{y}}+p\hat{\mathbf{z}}\right)e^{iq_{-}z}\right) & (z<0)
\end{array}\right.\\
&\mathbf{B}_{\mathrm{i}}^{*}(\mathbf{r})=\left\{ \begin{array}{llc}
\frac{E_{0}}{\omega_{\mathrm{i}}}e^{ipy}\left(r_{\mathrm{i}}\left(q_{\mathrm{i}}\hat{\mathbf{y}}+p\hat{\mathbf{z}}\right)e^{-iq_{\mathrm{i}}z}\right) &  & (0<z)\\
\frac{E_{0}}{\omega_{\mathrm{i}}}e^{ipy}\left(t_{\mathrm{i}1+}\left(-q_{+}\hat{\mathbf{y}}+p\hat{\mathbf{z}}\right)e^{iq_{+}z}+t_{\mathrm{i}2+}\left(q_{+}\hat{\mathbf{y}}+p\hat{\mathbf{z}}\right)e^{-iq_{+}z}+t_{\mathrm{i}1-}\left(-q_{-}\hat{\mathbf{y}}+p\hat{\mathbf{z}}\right)e^{iq_{-}z}+t_{\mathrm{i}2-}\left(q_{-}\hat{\mathbf{y}}+p\hat{\mathbf{z}}\right)e^{iq_{-}z}\right) &  & (z<0)
\end{array}\right.
\end{align}

We now utilize the Floquet eigenmodes discussed in the previous section, to link the signal and idler components in the bulk, $E_{\text{i}x}^\pm = \xi_{\text{i}x}^\pm E_{\text{s}x}^\pm$. 

Finally, we satisfy the boundary conditions at the interfaces to derive a set of 6 equations.

Air-hBN interface at $z=0$,
\begin{align} 
&1+r_{\mathrm{s}}=t_{\mathrm{s}1+}+t_{\mathrm{s}2+}+t_{\mathrm{s}1-}+t_{\mathrm{s}2-}\\
&r_{\mathrm{i}}=t_{\mathrm{s}1+}\xi_{\mathrm{i}1+}+t_{\mathrm{s}2+}\xi_{\mathrm{i}2+}+t_{\mathrm{s}1-}\xi_{\mathrm{i}1-}+t_{\mathrm{s}2-}\xi_{\mathrm{i}2-}\\
&-\frac{q_{\mathrm{s}}}{\omega_{\mathrm{s}}}\left(1-r_{\mathrm{s}}\right)=\frac{t_{\mathrm{s}1+}q_{+}}{\omega_{\mathrm{s}}}-\frac{t_{\mathrm{s}2+}q_{+}}{\omega_{\mathrm{s}}}+\frac{t_{\mathrm{s}1-}q_{-}}{\omega_{\mathrm{s}}}-\frac{t_{\mathrm{s}2-}q_{-}}{\omega_{\mathrm{s}}}\\
&\frac{q_{\mathrm{i}}}{\omega_{\mathrm{i}}}r_{\mathrm{i}}=-\frac{t_{\mathrm{s}1+}q_{+}\xi_{\mathrm{i}1+}}{\omega_{\mathrm{i}}}+\frac{t_{\mathrm{s}2+}q_{+}\xi_{\mathrm{i}2+}}{\omega_{\mathrm{i}}}-\frac{t_{\mathrm{s}1-}q_{-}\xi_{\mathrm{i}1-}}{\omega_{\mathrm{i}}}+\frac{t_{\mathrm{s}2-}q_{-}\xi_{\mathrm{i}2-}}{\omega_{\mathrm{i}}}
\end{align}

hBN-metal boundary at $z=-L$,
\begin{align} 
    &t_{\text{s}1+}e^{-iq_{+}L}+t_{\text{s}2+}e^{iq_{+}L}+t_{\text{s}1-}e^{-iq_{-}L}+t_{\text{s}2-}e^{iq_{-}L}=0\\
    &t_{\mathrm{s}1+}\xi_{\mathrm{i}1+}e^{-iq_{+}L}+t_{\mathrm{s}2+}\xi_{\mathrm{i}2+}e^{iq_{+}L}+t_{\mathrm{s}1-}\xi_{\mathrm{i}1-}e^{-iq_{-}L}+t_{\mathrm{s}2-}\xi_{\mathrm{i}2-}e^{iq_{-}L}=0
\end{align}
\subsection{p- polarized light}

For p- polarized incident light the magnetic field is oriented along the surface plane, particularly along the $x$ axis. Let us express the magnetic field in the air and within the hBN explicitly by introducing reflection and transmission coefficients,

\begin{align} 
 & \mathbf{B}_{\mathrm{s}}(\mathbf{r})=\left\{ \begin{array}{llc}
B_{0}\hat{\mathbf{x}}e^{ipy}\left(e^{-iq_{\mathrm{s}}z}+r_{\mathrm{s}}e^{iq_{\mathrm{s}}z}\right) &  & (0<z)\\
B_{0}\hat{\mathbf{x}}e^{ipy}\left(t_{\text{s}1+}e^{iq_{+}z}+t_{\text{s}2+}e^{-iq_{+}z}+t_{\text{s}1-}e^{iq_{-}z}+t_{\text{s}2-}e^{-iq_{-}z}\right) &  & (z<0)
\end{array}\right.\\
& \mathbf{B}_{\mathrm{i}}^{*}(\mathbf{r})=\left\{ \begin{array}{llc}
B_{0}\hat{\mathbf{x}}e^{ipy}\left(r_{\mathrm{i}}e^{-iq_{\mathrm{i}}z}\right) &  & (0<z)\\
B_{0}\hat{\mathbf{x}}e^{ipy}\left(t_{\text{i}1+}e^{iq_{+}z}+t_{\text{i}2+}e^{-iq_{+}z}+t_{\text{i}1-}e^{iq_{-}z}+t_{\text{i}2-}e^{-iq_{-}z}\right) &  & (z<0)
\end{array}\right.
\end{align}

We now use these magnetic fields and Maxwell equations with Floquet modes to recover the corresponding electric fields and link the idler transmission coefficients to the signal ones. Let us start with the signal component.

In the air the procedure is straightforward. The electric and magnetic fields
are perpendicular to one another, and obey the Maxwell's equation \eqref{eqSup:MaxSup}. 
For the signal component in the air this yields, 
\begin{align} 
 & \mathbf{E}_{\mathrm{s}}(\mathbf{r})=c^{2}\frac{B_{0}}{\omega_{\mathrm{s}}}e^{ipy}\left[e^{-iq_{\mathrm{s}}z}\left(q_{\mathrm{s}}\hat{\mathbf{y}}+p\hat{\mathbf{z}}\right)+r_{\mathrm{s}}e^{iq_{\mathrm{s}}z}\left(-q_{\mathrm{s}}\hat{\mathbf{y}}+p\hat{\mathbf{z}}\right)\right]\qquad(0<z)
\end{align}

The analysis is more complicated in the bulk.  We first need to satisfy the Maxwell equation; and since they are to be satisfied for arbitrary $z$, we can consider each term in the magnetic field expression individually. 

Let us start with the term containing $t_{\text{s}1+}$. The electric field should be of a form,
\begin{align*} 
 & \mathbf{E}_{\mathrm{s},1}^{+}(\mathbf{r})= e^{ipy}e^{iq_{+}z}\left(E_{sy}^{+}\hat{\mathbf{y}}+E_{sz}^{+}\hat{\mathbf{z}}\right)
\end{align*}
We need to satisfy not only the Maxwell equations but the linearized
driven equations as well. That is why we express explicitly the $z$ component through the $y$ component, $E_{sz}^{+}=\xi_{sz}^{+}E_{sy}^{+}$.
Thus, the electric field is written as
\begin{align} 
 & \mathbf{E}_{\mathrm{s},1}^{+}(\mathbf{r})=e^{ipy}e^{iq_{+}z}E_{sy}^{+}\left(\hat{\mathbf{y}}+\xi_{sz}^{+}\hat{\mathbf{z}}\right)
\end{align}
To employ the aforementioned Maxwell equation, we calculate the curl of the electric field from above
\begin{align} 
 & \nabla\times\mathbf{E}_{\mathrm{s},1}^{+}(\mathbf{r})=E_{sy}^{+}e^{ipy}e^{iq_{+}z}i\left(\xi_{sz}^{+}p-q_{+}\right)\hat{\mathbf{x}}
\end{align}
The magnetic field is, on one hand, linked to the electric field through the Maxwell's equation; and, on the other hand, has the form we started from. Thus, the two expressions should be identical, 
\begin{align} 
& \textbf{B} =-i\frac{\nabla\times\textbf{E}}{\omega_\text{s}}=\frac{1}{\omega_\text{s}}E_{\text{s}y}^{+}e^{ipy}e^{iq_{+}z}\left(\xi_{\text{s}z}^{+}p-q_{+}\right)\hat{\mathbf{x}}\equiv B_{0}t_{\text{s}1+}e^{ipy}e^{iq_{+}z}\hat{\mathbf{x}}
\end{align}
Therefore, we can express the electric field $E_{\text{s}y}^{+}$ as
\begin{align} 
    E_{\text{s}y}^{+}=\frac{B_{0}\omega_{\text{s}}t_{s1+}}{\xi_{\text{s}z}^{+}p-q_{+}}
\end{align}
By repeating the procedure for the remaining signal terms, and taking into account $\xi_{sz}^{+}\left(-q_{+}\right)=-\xi_{\text{s}z}^{+}\left(q_{+}\right)$ we arrive to the following expression for the signal component of electric field in the bulk,
\begin{align} 
 & B_{0}\omega_{\text{s}}e^{ipy}\left(\frac{t_{\text{s}1+}\left(\hat{\mathbf{y}}+\xi_{\text{s}z}^{+}\hat{\mathbf{z}}\right)}{\xi_{\text{s}z}^{+}p-q_{+}}e^{iq_{+}z}+\frac{t_{\text{s}2+}\left(\hat{\mathbf{y}}-\xi_{\text{s}z}^{+}\hat{\mathbf{z}}\right)}{-\xi_{\text{s}z}^{+}p+q_{+}}e^{-iq_{+}z}+\frac{t_{\text{s}1-}\left(\hat{\mathbf{y}}+\xi_{\text{s}z}^{-}\hat{\mathbf{z}}\right)}{\xi_{\text{s}z}^{-}p-q_{-}}e^{iq_{-}z}+\frac{t_{\text{s}2-}\left(\hat{\mathbf{y}}-\xi_{\text{s}z}^{-}\hat{\mathbf{z}}\right)}{-\xi_{\text{s}z}^{-}p+q_{-}}e^{-iq_{-}z}\right)
\end{align}

We also derive similar expressions for the idler component. The only difference is that we must once again use the Floquet eigenmodes to express $t_{\text{i}\pm}$ via $t_{\text{s}\pm}$. To do so, we express $E_{\text{i}y}^{*+}=\xi_{\text{i}y}^{+}E_{\text{s}y}^{+}$ and
$E_{\text{i}z}^{*+}=\xi_{\text{i}z}^{+}E_{\text{s}y}^{+}$, allowing us to preserve
only $E_{\text{s}y}^{+}$. Then the idler electric field is easily derived
from the signal electric field found above. We start with the first term,
\begin{align} 
\mathbf{E}_{\mathrm{i}1}^{*}(\mathbf{r}) & =B_{0}\omega_{s}e^{ipy}\frac{t_{s1+}\left(\xi_{iy}^{+}\hat{\mathbf{y}}+\xi_{iz}^{+}\hat{\mathbf{z}}\right)}{\xi_{sz}^{+}p-q_{+}}e^{iq_{+}z}
\end{align}
From this we find the corresponding magnetic field:
\begin{align} 
&\textbf{B}=i\frac{\nabla\times\textbf{E}}{\omega_\text{i}}=-\frac{\omega_{\mathrm{s}}}{\omega_{\mathrm{i}}}B_{0}e^{ipy}e^{iq_{+}z}\frac{t_{\text{s}1+}\left(\xi_{\text{i}z}^{+}p-\xi_{\text{i}y}^{+}q_{+}\right)}{\xi_{\text{s}z}^{+}p-q_{+}}\hat{\mathbf{x}} \equiv B_{0}t_{\mathrm{i1}+}e^{ipy}e^{iq_{+}z}\hat{\mathbf{x}}\\
&\Rightarrow t_{\mathrm{i1}+}=-\frac{\omega_{\mathrm{s}}}{\omega_{\mathrm{i}}}\frac{t_{\mathrm{s}1+}\left(\xi_{\text{i}z}^{+}p-\xi_{\text{i}y}^{+}q_{+}\right)}{\xi_{\text{s}z}^{+}p-q_{+}}
\end{align}

The other elements are recovered similarly, taking into account the simple geometric observation that $\xi_{iz}^{+}\left(-q_{+}\right)=-\xi_{iz}^{+}\left(q_{+}\right)$, however, $\xi_{iy}^{+}\left(-q_{+}\right)=\xi_{iy}^{+}\left(q_{+}\right)$.

Combining all the findings we obtain the fields everywhere. For the magnetic field,
\begin{align} 
 & \mathbf{B}_{\mathrm{s}}(\mathbf{r})=\left\{ \begin{array}{llc}
B_{0}\hat{\mathbf{x}}e^{ipy}\left(e^{-iq_{\mathrm{s}}z}+r_{\mathrm{s}}e^{iq_{\mathrm{s}}z}\right) &  & (0<z)\\
B_{0}\hat{\mathbf{x}}e^{ipy}\left(t_{\text{s}1+}e^{iq_{+}z}+t_{\text{s}2+}e^{-iq_{+}z}+t_{\text{s}1-}e^{iq_{-}z}+t_{\text{s}2-}e^{-iq_{-}z}\right) &  & (z<0)
\end{array}\right.\\
& \mathbf{B}_{\mathrm{i}}^{*}(\mathbf{r})=\left\{ \begin{array}{llc}
B_{0}\hat{\mathbf{x}}e^{ipy}\left(r_{\mathrm{i}}e^{-iq_{\mathrm{i}}z}\right)&&(0<z)\\-B_{0}\frac{\omega_{\mathrm{s}}}{\omega_{\mathrm{i}}}\hat{\mathbf{x}}e^{ipy}\left(\frac{\xi_{iz}^{+}p-\xi_{iy}^{+}q_{+}}{\xi_{sz}^{+}p-q_{+}}t_{s1+}e^{iq_{+}z}+\frac{\xi_{iz}^{+}p-\xi_{iy}^{+}q_{+}}{\xi_{sz}^{+}p-q_{+}}t_{s2+}e^{-iq_{+}z}+\frac{\xi_{iz}^{-}p-\xi_{iy}^{-}q_{-}}{\xi_{sz}^{-}p-q_{-}}t_{s1-}e^{iq_{-}z}+\frac{\xi_{iz}^{-}p-\xi_{iy}^{-}q_{-}}{\xi_{sz}^{-}p-q_{-}}t_{s2-}e^{-iq_{-}z}\right)&&(z<0)
\end{array}\right.
\end{align}
For the electric field,
\begin{align} 
 & \mathbf{E}_{\mathrm{s}}(\mathbf{r})=\left\{ \begin{array}{llc}
c^{2}\frac{B_{0}}{\omega_{\mathrm{s}}}e^{ipy}\left[e^{-iq_{\mathrm{s}}z}\left(q_{\mathrm{s}}\hat{\mathbf{y}}+p\hat{\mathbf{z}}\right)+r_{\mathrm{s}}e^{iq_{\mathrm{s}}z}\left(-q_{\mathrm{s}}\hat{\mathbf{y}}+p\hat{\mathbf{z}}\right)\right]&&(0<z)\\B_{0}\omega_{s}e^{ipy}\left(\frac{t_{s1+}\left(\hat{\mathbf{y}}+\xi_{sz}^{+}\hat{\mathbf{z}}\right)}{\xi_{sz}^{+}p-q_{+}}e^{iq_{+}z}+\frac{t_{s2+}\left(\hat{\mathbf{y}}-\xi_{sz}^{+}\hat{\mathbf{z}}\right)}{-\xi_{sz}^{+}p+q_{+}}e^{-iq_{+}z}+\frac{t_{s1-}\left(\hat{\mathbf{y}}-\xi_{sz}^{-}\hat{\mathbf{z}}\right)}{-\xi_{sz}^{-}p-q_{-}}e^{iq_{-}z}+\frac{t_{s2-}\left(\hat{\mathbf{y}}-\xi_{sz}^{-}\hat{\mathbf{z}}\right)}{-\xi_{sz}^{-}p+q_{-}}e^{-iq_{-}z}\right)&&(z<0)
\end{array}\right.\\
& \mathbf{E}_{\mathrm{i}}^{*}(\mathbf{r})=\left\{ \begin{array}{llc}
-c^{2}\frac{B_{0}}{\omega_{\mathrm{i}}}e^{ipy}r_{\mathrm{i}}e^{-iq_{\mathrm{i}}z}\left(q_{\mathrm{i}}\hat{\mathbf{y}}+p\hat{\mathbf{z}}\right)&&(0<z)\\B_{0}\omega_{s}e^{ipy}\left(\frac{t_{s1+}\left(\xi_{iy}^{+}\hat{\mathbf{y}}+\xi_{iz}^{+}\hat{\mathbf{z}}\right)}{\xi_{sz}^{+}p-q_{+}}e^{iq_{+}z}+\frac{t_{s2+}\left(\xi_{iy}^{+}\hat{\mathbf{y}}-\xi_{iz}^{+}\hat{\mathbf{z}}\right)}{-\xi_{sz}^{+}p+q_{+}}e^{-iq_{+}z}+\frac{t_{s1-}\left(\xi_{iy}^{-}\hat{\mathbf{y}}-\xi_{iz}^{-}\hat{\mathbf{z}}\right)}{-\xi_{sz}^{-}p-q_{-}}e^{iq_{-}z}+\frac{t_{s2-}\left(\xi_{iy}^{-}\hat{\mathbf{y}}-\xi_{iz}^{-}\hat{\mathbf{z}}\right)}{-\xi_{sz}^{-}p+q_{-}}e^{-iq_{-}z}\right)&&(z<0)
\end{array}\right.
\end{align}

Using the boundary conditions at the interfaces, we derive a set of 6 equations: 

Air-hBN boundary at $z=0$:
\begin{align} 
&1+r_{s}=t_{s1+}+t_{s2+}+t_{s1-}+t_{s2-}\\
&r_{i}=-\frac{\omega_{\mathrm{s}}}{\omega_{\mathrm{i}}}\left(\frac{\xi_{iz}^{+}p-\xi_{iy}^{+}q_{+}}{\xi_{sz}^{+}p-q_{+}}t_{s1+}+\frac{\xi_{iz}^{+}p-\xi_{iy}^{+}q_{+}}{\xi_{sz}^{+}p-q_{+}}t_{s2+}+\frac{\xi_{iz}^{-}p-\xi_{iy}^{-}q_{-}}{\xi_{sz}^{-}p-q_{-}}t_{s1-}+\frac{\xi_{iz}^{-}p-\xi_{iy}^{-}q_{-}}{\xi_{sz}^{-}p-q_{-}}t_{s2-}\right)\\
&c^{2}\frac{q_{\mathrm{s}}}{\omega_{\mathrm{s}}}\left(1-r_{\mathrm{s}}\right)=\omega_{s}\left(\frac{t_{s1+}}{\xi_{sz}^{+}p-q_{+}}-\frac{t_{s2+}}{\xi_{sz}^{+}p-q_{+}}+\frac{t_{s1-}}{\xi_{sz}^{-}p-q_{-}}-\frac{t_{s2-}}{\xi_{sz}^{-}p-q_{-}}\right)\\
&-c^{2}\frac{q_{i}}{\omega_{\mathrm{i}}}r_{\mathrm{i}}=\omega_{s}\left(\frac{t_{s1+}\xi_{iy}^{+}}{\xi_{sz}^{+}p-q_{+}}-\frac{t_{s2+}\xi_{iy}^{+}}{\xi_{sz}^{+}p-q_{+}}+\frac{t_{s1-}\xi_{iy}^{-}}{\xi_{sz}^{-}p-q_{-}}-\frac{t_{s2-}\xi_{iy}^{-}}{\xi_{sz}^{-}p-q_{-}}\right)
\end{align}

hBN-metal boundary at $z=-L$:
\begin{align} 
 & \frac{t_{s1+}}{\xi_{sz}^{+}p-q_{+}}e^{-iq_{+}L}-\frac{t_{s2+}}{\xi_{sz}^{+}p-q_{+}}e^{iq_{+}L}+\frac{t_{s1-}}{\xi_{sz}^{-}p-q_{-}}e^{-iq_{-}L}-\frac{t_{s2-}}{\xi_{sz}^{-}p-q_{-}}e^{iq_{-}L}=0\\
& \frac{t_{s1+}\xi_{iy}^{+}}{\xi_{sz}^{+}p-q_{+}}e^{-iq_{+}L}-\frac{t_{s2+}\xi_{iy}^{+}}{\xi_{sz}^{+}p-q_{+}}e^{iq_{+}L}+\frac{t_{s1-}\xi_{iy}^{-}}{\xi_{sz}^{-}p-q_{-}}e^{-iq_{-}L}-\frac{t_{s2-}\xi_{iy}^{-}}{\xi_{sz}^{-}p-q_{-}}e^{iq_{-}L}=0
\end{align}

The above equation may obtain simpler forms if we redefine the transition coefficient as $\frac{t_{s1+}}{\xi_{sz}^{+}p-q_{+}}\frac{\omega_{s}}{c}\rightarrow t_{s1+}$ and $\frac{t_{i}}{\xi_{iz}^{0}p-q_{i0}}\frac{\omega_{i}}{c}\rightarrow t_{i}$. However, since the equations can only be solved numerically we will employ the ones from above.  

\section{\label{SecSup:2THz} Characteristics of the 2THz instability}

In the main text we discussed some of the important properties of the 2 Thz instability. In this section we briefly outline the role of angle and pumping intensity in the poles' structure. 

The Fig. \ref{figSup:2Thz_prop} summarizes the main results. As discussed in the main text, the increase in the slab thickness pushes the poles towards the real axis. Eventually, they are shifted to the upper half-plane, leading to lasing. 

The increasing angle of incident has a rather similar effect. When shining the probe pulse at a finite angle of incidence, it creates Floquet eigenmodes within hBN that also propagate at a finite angle. This effectively lengthens the optical path. This effect closely resembles the increase in slab thickness, leading to similar outcomes as indicated in the Fig. \ref{figSup:2Thz_prop}b).

Finally, the increasing pumping shifts the resonant region to larger frequencies and increases its width. Simultaneously, the poles are again pushed towards the real axis with a subsequent expansion into the upper half-plane. This is depicted in the Fig. \ref{figSup:2Thz_prop}c). 

\begin{figure*}[h!]
\begin{centering}
\includegraphics[width=0.9\textwidth]{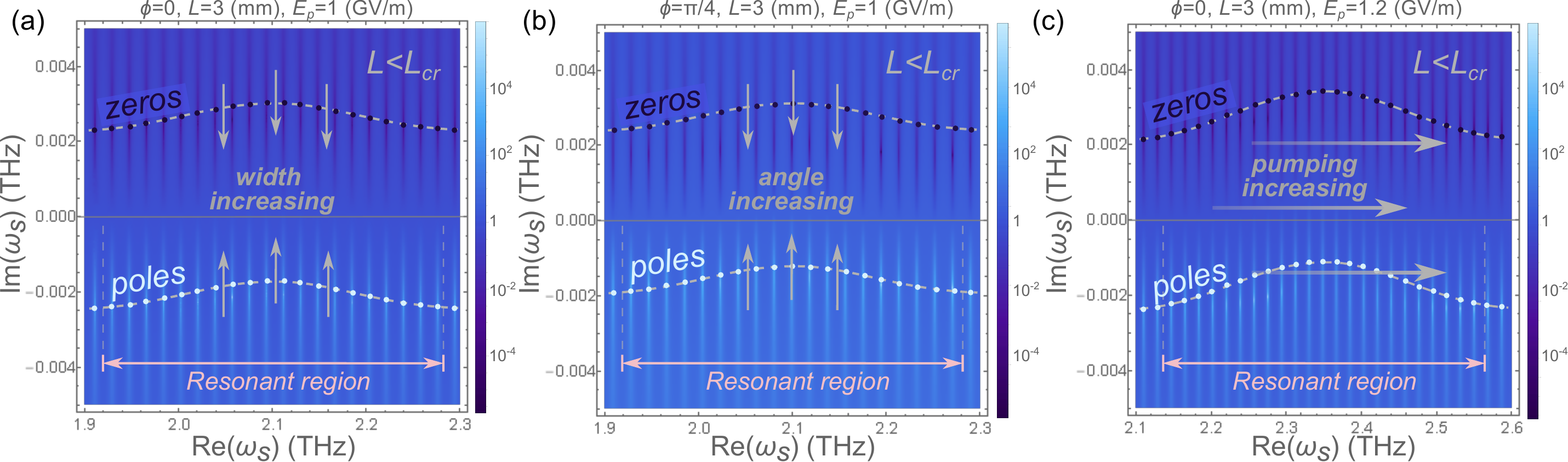}
\par\end{centering}
\caption{\label{figSup:2Thz_prop} Poles’ structure of reflectivity (in log scale) around $\omega_\text{s} = 2.1$ THz instability. (a) Showing the effects of increasing width discussed in the main text. (b) The influence of the increasing angle of incidence. Finite angle lengthens the waves' optical paths inside the slab, thus is effectively equivalent to the effects of increasing width. (c) The effects of increasing pumping. The center of the resonant region is shifted to larger frequencies (see Fig. \ref{fig:Complex_Phase_Diagram}a) in the main text), and the poles are pushed towards the real axis}
\end{figure*}

\section{Advanced cavity designs \label{SecSup:Cavity} }
\subsection{Suppressing the 24-27 THz instabilities with metal frame}

The main text highlights that the instabilities in the $24$-$27$ THz range are dominant and could potentially mask the effects of the $2$ THz instability. One strategy to accentuate the latter involves the design of more sophisticated cavities than those primarily discussed.

As one possibility, we propose leveraging the skin effect as a potential solution. Signals in the $24$-$27$ THz range have frequencies an order of magnitude higher than the $2$ THz signal. Given that the penetration depth is inversely proportional to the square root of the frequency — $\delta(\omega) = \sqrt{2\rho/ \omega \mu_0 \mu}\sim \omega^{-1/2}$ — the depth varies substantially between the two frequencies. Here, $\rho$ and $\mu$ are the resistance and the relative permeability of the metal, respectively. Specifically, in the case of copper, $\delta(2\text{ THz}) \approx 50$ nm, whereas $\delta(25\text{ THz}) \approx 15$ nm. By encasing the hBN in a thin, approximately $25$ nm, layer of metal, we can make the material nearly transparent to the lower-frequency signal while fully reflecting the higher-frequency one. This would confine the $24$-$27$ THz pulses within the material, preventing energy depletion.

\subsection{Increasing the effective thickness}

As we discuss in the main text, some of the effects described might necessitate  crystals of a large thickness to observe any amplification of the initial signal. 

To avoid the challenging process of producing crystals of large sizes we can slightly alter the experimental setup. One straightforward solution is to add an additional metal layer above the hBN slab, slightly smaller than the slab itself, as depicted in Figure \ref{figSup:Modified_setup}. By directing the probe pulse at the hBN, the transmitted wave will undergo multiple reflections before being emitted back into the air. This effectively lengthens the optical path of the transmitted wave without necessitating an increase in the physical thickness of the slab.

\begin{figure}[h!]
\begin{centering}
\includegraphics[width=0.3\columnwidth]{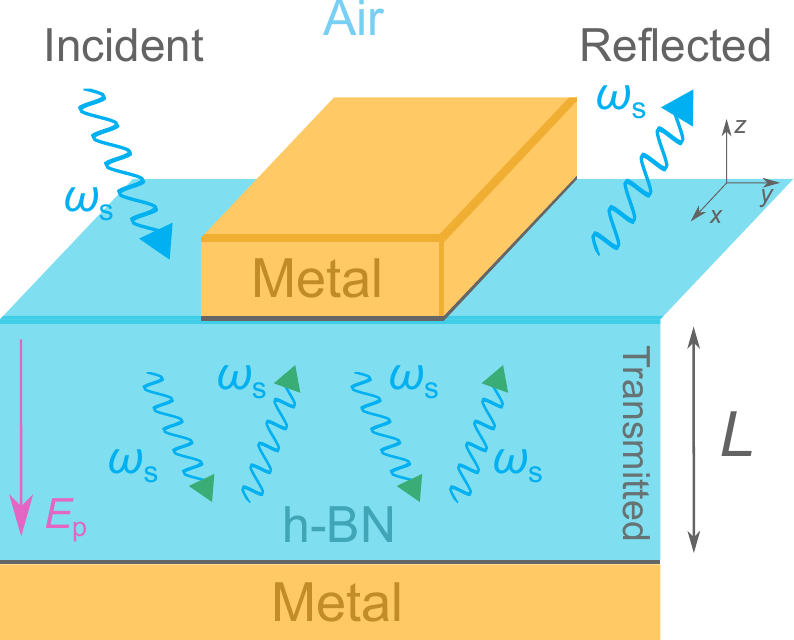}
\par\end{centering}
\caption{ Schematic representation of the adapted experimental setup, featuring an added metallic layer atop hBN slab's surface. The layer is strategically positioned to not fully cover the top surface, enabling the probe pulse to directly interact with the hBN. Light that penetrates the slab subsequently undergoes multiple reflections between the two metallic layers before being re-emitted into the air on the opposite side. This configuration effectively extends the optical path of the wave within the bulk material without requiring an increase in the slab's actual thickness.
\label{figSup:Modified_setup}}
\end{figure}

\end{widetext}
\end{document}